\documentclass[11pt]{article}
\pdfoutput=1
\usepackage[utf8]{inputenc}
\usepackage{jheppub}
\usepackage{subcaption}
\usepackage{xparse}

\NewDocumentCommand{\binomial}{omm}
 {%
  \genfrac(){0pt}{}{#2}{#3}%
  \IfValueT{#1}{_{\!#1}}%
 }
\NewDocumentCommand{\eulerian}{omm}
 {%
  \genfrac<>{0pt}{}{#2}{#3}%
  \IfValueT{#1}{_{\!#1}}%
 }

\newcommand{\nn}{\nonumber}
\newcommand{\doe}{\partial}
\newcommand{\be}{\begin{equation}}
\newcommand{\ee}{\end{equation}}
\newcommand{\bea}{\begin{eqnarray}}
\newcommand{\eea}{\end{eqnarray}}
\newcommand{\bdm}{\begin{displaymath}}
\newcommand{\edm}{\end{displaymath}}
\newcommand{\bse}{\begin{subequations}}
\newcommand{\ese}{\end{subequations}}

\newcommand{\mr}{\mathrm}

\newcommand{\mc}{\mathcal}

\newcommand{\bs}{\boldsymbol}

\usepackage{latexsym}

\def\ie{i.e. }
\def\eg{e.g. }
\def\etc{etc. }

\def\eqn#1{eq.~\eqref{#1}}
\def\eqns#1#2{eqs.~\eqref{#1} and~\eqref{#2}}

\def\fig#1{figure~{\ref{#1}}}

\def\sec#1{section~{\ref{#1}}}

\def\app#1{appendix~{\ref{#1}}}

\def\bra#1{\langle #1|}
\def\ket#1{|#1 \rangle}
\def\la{\lambda}
\def\lb{\tilde{\lambda}}
\def\braket#1{\langle #1 \rangle}

\def \be {\begin{equation}}
\def \ee {\end{equation}}
\def \beal {\begin{equation}\begin{aligned}}
\def \eeal {\end{aligned}\end{equation}}

\begin{document}

\title{Scattering of Spinning Black Holes from Exponentiated Soft Factors}
\author[a,b,c]{Alfredo Guevara,}
\author[d]{Alexander Ochirov,}
\author[e]{and Justin Vines}

\affiliation[a]{Perimeter Institute for Theoretical Physics, Waterloo, ON N2L 2Y5, Canada}
\affiliation[b]{Department of Physics $\&$ Astronomy, University of Waterloo, Waterloo, ON N2L 3G1, Canada}
\affiliation[c]{CECs Valdivia \& Departamento de F\'isica, Universidad de Concepci\'on, Casilla 160-C,\\ Concepci\'on, Chile}
\affiliation[d]{ETH Z\"urich, Institut f\"ur Theoretische Physik,
Wolfgang-Pauli-Str. 27, 8093 Z\"urich, Switzerland}
\affiliation[e]{Max Planck Institute for Gravitational Physics (Albert Einstein Institute), Am M\"uhlenberg 1, Potsdam 14476, Germany}

\emailAdd{aguevara@pitp.ca,aochirov@phys.ethz.ch,justin.vines@aei.mpg.de}

\abstract{We provide evidence that the classical scattering of two spinning black holes is controlled by the soft expansion of exchanged gravitons.  We show how an exponentiation of Cachazo-Strominger soft factors, acting on massive higher-spin amplitudes, can be used to find spin contributions to the aligned-spin scattering angle, conjecturally extending previously known results to higher orders in spin at one-loop order.  The extraction of the classical limit is accomplished via the on-shell leading-singularity method and using massive spinor-helicity variables.  The three-point amplitude for arbitrary-spin massive particles minimally coupled to gravity is expressed in an exponential form, and in the infinite-spin limit it matches the effective stress-energy tensor of the linearized Kerr solution.  A four-point gravitational Compton amplitude is obtained from an extrapolated soft theorem, equivalent to gluing two exponential three-point amplitudes, and becomes itself an exponential operator.  The construction uses these amplitudes to: 1) recover the known tree-level scattering angle at all orders in spin, 2) recover the known one-loop linear-in-spin interaction, 3) match a previous conjectural expression for the one-loop scattering angle at quadratic order in spin, 4) propose new one-loop results through quartic order in spin.  These connections link the computation of higher-multipole interactions to the study of deeper orders in the soft expansion.}

\maketitle
\addtocontents{toc}{\protect\setcounter{tocdepth}{1}}

\section{Introduction}
\label{sec:intro}

In 2014 Cachazo and Strominger~\cite{Cachazo:2014fwa} showed that the soft limit of tree-level gravity amplitudes is controlled by the action of the angular momentum operator $J^{\mu \nu}$, i.e.

\be
   {\cal M}_{n+1} = \sum_{i=1}^n
   \left[ \frac{(p_i\cdot\varepsilon)^2}{p_i\cdot k}
        +i\frac{(p_i\cdot\varepsilon)
                (k_\mu \varepsilon_\nu J_i^{\mu\nu})}{p_i\cdot k}
        - \frac{1}{2}\frac{(k_\mu \varepsilon_\nu J_i^{\mu\nu})^2}
                          {p_i\cdot k}
   \right] {\cal M}_n + {\cal O}(k^2) ,
\label{eq:cachazostrominger}
\ee
up to sub-subleading order. Here the soft momentum $k$ corresponds to the external soft graviton, and we have constructed its polarization tensor as
$\varepsilon_{\mu\nu}=\varepsilon_{\mu}\varepsilon_{\nu}$.
The sum is over the remaining external particles with momenta $p_i^\mu$,
and the operators $J_i^{\mu \nu}$ acting on them include both orbital and spin parts of the angular momentum. The first term is simply
the standard Weinberg soft factor~\cite{Weinberg:1965nx}, whose universality is associated to the equivalence principle.
Following the QED results of Low \cite{Low:1954kd,Low:1958sn}, the subleading behaviour of gravity amplitudes was first studied long ago by Gross and Jackiw \cite{Gross:1968in,Jackiw:1968zza}. Indeed, it was already observed in \cite{Gross:1968in,Jackiw:1968zza} that the subleading soft theorem follows from gauge invariance (see \cite{White:2011yy,Bern:2014vva} for a modern perspective), and because of this, it also adopts a universal form up to subleading order. Starting at sub-subleading order the soft expansion can depend on the matter content and EFT operators present in the theory \cite{Laddha:2017ygw,Sen:2017xjn,Bianchi:2014gla}, although it is known that gauge invariance still provides partial information at all orders \cite{Hamada:2018vrw,Li:2018gnc}. On a different front, the realization that soft theorems correspond to Ward identities
for asymptotic symmetries at null infinity \cite{Strominger:2013jfa}
has led to impressive and wide-reaching developments~\cite{He:2014laa,Cachazo:2014fwa,Cachazo:2014dia,Kapec:2014opa,Bern:2014vva,Dumitrescu:2015fej,Campiglia:2016hvg},
see \cite{Strominger:2017zoo} for a recent review. Following such correspondence, an infinite tower of Ward identities has indeed been proposed to follow from all orders in the soft expansion \cite{Campiglia:2018dyi}.

\begin{figure}[t]
\centering
\subcaptionbox{\label{fig:intro1}}{
\includegraphics[width=0.45\textwidth]{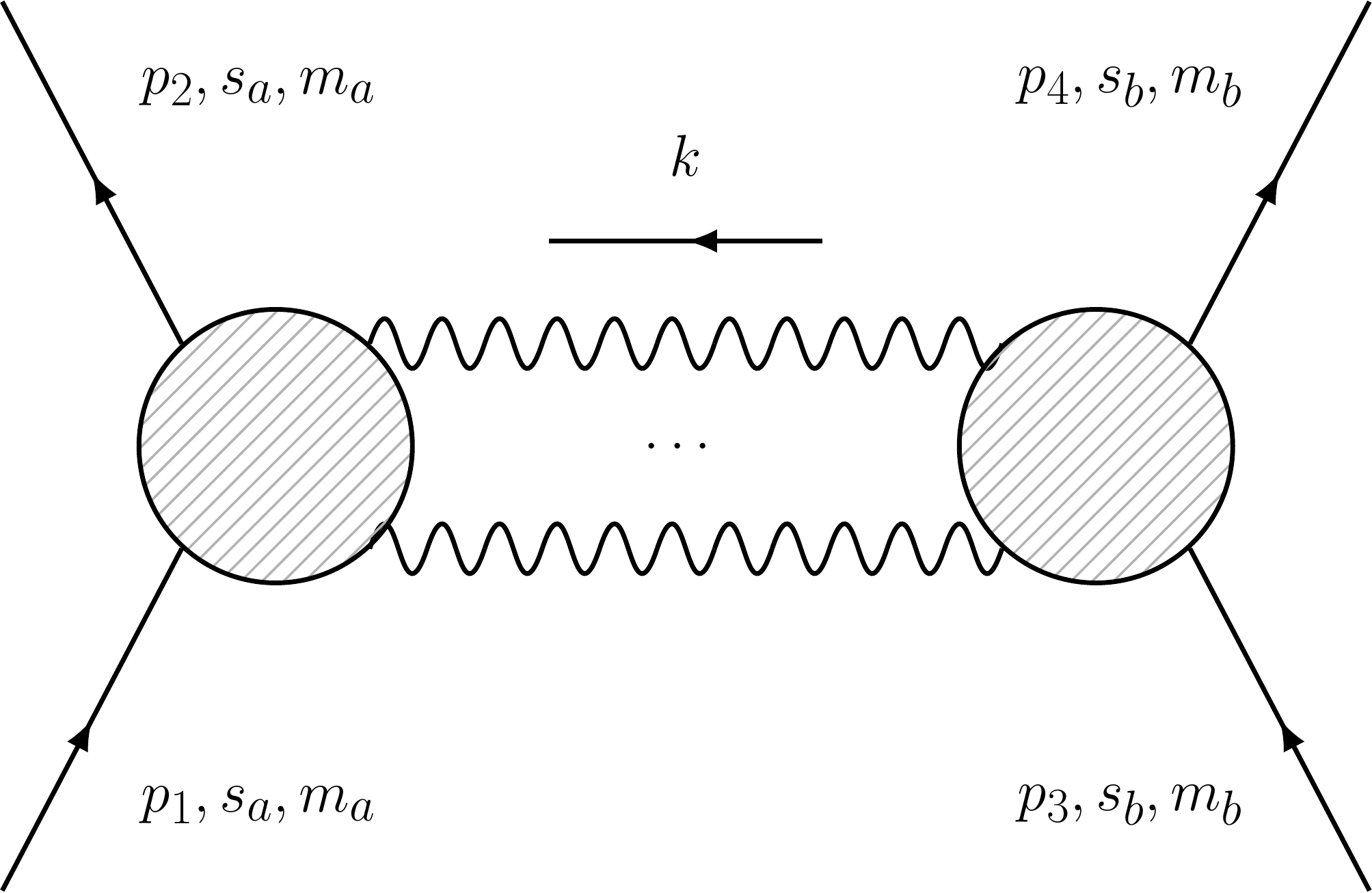} 
\vspace{12.5pt}
} \hfill
\subcaptionbox{\label{fig:intro2}}{
\includegraphics[width=0.45\textwidth]{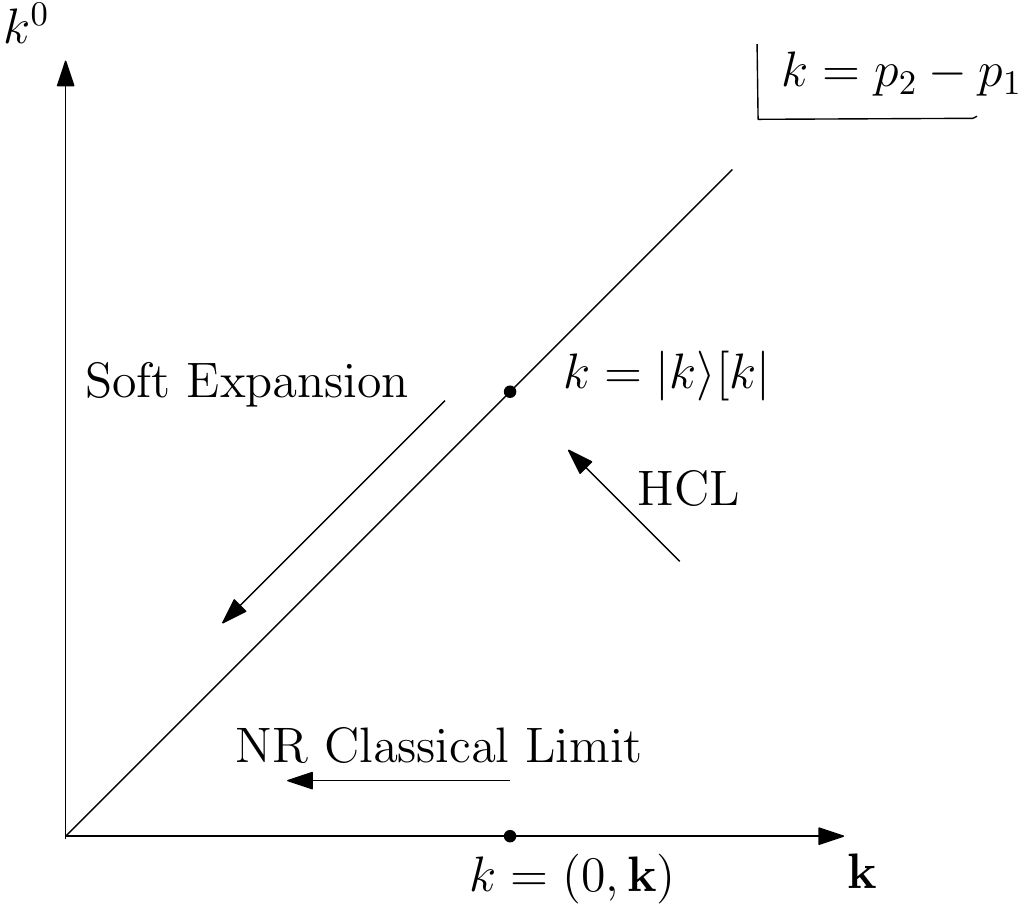} 
}
\caption{\textbf{(a)} Four-point amplitude involving the exchange of soft gravitons, which leads to classical observables. The external massive states are interpreted as two black-hole sources.\\
\textbf{(b)} Comparison between the HCL and the non-relativistic limit in the COM frame \cite{Holstein:2008sx,Holstein:2008sw,Vaidya:2014kza}. Spin effects require subleading orders in the nonrelativistic (NR) classical limit, but can be fully determined at the leading order in HCL through the soft expansion.}
\end{figure}

Recently, a classical version of the soft theorem up to sub-subleading order
has been used by Laddha and Sen~\cite{Laddha:2018rle}
to derive the spectrum of the radiated power
in black-hole scattering with external soft graviton insertions.
This relies on the remarkable fact that
conservative and non-conservative long-range effects
of interacting black holes can be computed from the
scattering of massive point-like sources~\cite{Duff:1973zz,
BjerrumBohr:2002ks,Neill:2013wsa,Bjerrum-Bohr:2018xdl}.
Indeed, rotating black holes can be treated via a spin-multipole expansion, the order $2s$ of which can be reproduced
by scattering spin-$s$ minimally coupled particles
exchanging gravitons~\cite{Vaidya:2014kza},
as illustrated in \fig{fig:intro1}. The matching between these amplitudes with spin and a non-relativistic potential for black-hole scattering has been performed explicitly in the post-Newtonian (PN) framework \cite{Vaidya:2014kza,Holstein:2008sx,Guevara:2017csg}.

Here we present a complementary picture to the one of \cite{Laddha:2018rle}
by employing the soft theorem in the conservative sector (\ie no external gravitons), focusing
on rotating black holes and at the same time extending the soft factor in \eqref{eq:cachazostrominger} to higher orders in the soft expansion. This is achieved in the following way: It was shown by one of the authors in~\cite{Guevara:2017csg}
that the classical ($\hbar$-independent) piece of the spin-$s$ amplitude
can be extracted from a covariant Holomorphic Classical Limit (HCL),
which sets the external kinematics
such that the momentum transfer $k$ between the massive sources is null.
On the support of the leading-singularity (LS) construction~\cite{Cachazo:2017jef}, which drops $\mathcal{O}(\hbar)$ parts,
the condition $k^2=0$ reduces the amplitude
to a purely classical expansion in spin multipoles of the form $\sim k^n S^n$,
where $S$ carries the intrinsic angular momentum of the black hole (see \fig{fig:intro2}).
This precisely matches the soft expansion
once the momentum transfer is recognized as the graviton momentum
and the classical spin vector $S$ is identified
with the angular momentum $J_i$ of the matter particles.

To see the soft expansion more explicitly, consider the energy-momentum tensor of a single linearized Kerr black hole, which has recently been written down in an exponential form by one of the authors~\cite{Vines:2017hyw}:
\be
T^{\mu\nu}(-k)
= 2\pi \delta{(p\cdot k)} p^{(\mu}\exp(a*ik)^{\nu)}{}_\rho\;\!p^\rho
+ \mathcal{O}(G),
\ee
where $(a*k)^\mu{}_\nu=\epsilon^\mu{}_{\nu\rho\sigma}a^\rho k^\sigma$, and $a^\mu={S^\mu}/{m}$ is the rescaled spin vector of the black hole.
The magnitude $a$ is exactly the radius of its ring singularity.
Here we have performed a Fourier transform
of the worldline formulas~(18) and~(32a) of~\cite{Vines:2017hyw}.
Now, the interaction vertex between a
graviton and a massive source corresponds to the contraction $-h_{\mu\nu}T^{\mu\nu}$.
After we take the graviton to be on-shell and replace $h_{\mu\nu}(k)$ by
$2\pi\delta{(k^2)}\varepsilon_\mu\varepsilon_\nu$, the vertex becomes
\be
   h_{\mu\nu}(k)T^{\mu\nu}(-k)
    = (2\pi)^2 \delta{(k^2)} \delta{(p\cdot k)}
      (p\cdot\varepsilon)\;\!\varepsilon_{\mu}p_{\nu}
      \bigg[ \eta^{\mu\nu} - i\epsilon^{\mu\nu\rho\sigma} k_\rho a_\sigma
           + \frac{1}{2} \eta^{\mu\nu} (a \cdot k)^2 + \mathcal{O}(k^3)
      \bigg] ,
\label{eq:tuvkerr3}
\ee
where we have used the support of the delta functions.
This expression can be written in a simple form
by introducing the spin tensor
\begin{equation}
   S^{\mu\nu}=\epsilon^{\mu\nu\rho\sigma} p_\rho a_\sigma
   \qquad \Rightarrow \qquad
   a_\lambda = \frac{1}{2m^2} \epsilon_{\la\mu\nu\rho} S^{\mu\nu} p^\rho ,
\label{eq:spinvector2spintensor}
\end{equation}
satisfying $S^{\mu\nu}p_\nu=0$, after which it becomes
\beal
   h_{\mu\nu}(k) T^{\mu\nu}(-k) &
    = (2\pi)^2 \delta{(k^2)} \delta{(p \cdot k)} (p\cdot\varepsilon)^2
      \exp\!\bigg(\!{-i}\frac{k_\mu \varepsilon_\nu S^{\mu\nu}}
                             {p\cdot\varepsilon}\bigg) \\ &
    = (2\pi)^2 \delta{(k^2)} \delta{(p \cdot k)} (p\cdot\varepsilon)^2
      \Bigg[ 1 - i \frac{k_\mu \varepsilon_\nu S^{\mu\nu}}{p\cdot\varepsilon}
               - \frac{1}{2} \bigg(\frac{k_\mu \varepsilon_\nu S^{\mu\nu}}
                                   {p\cdot\varepsilon}\bigg)^{\!2}
     + {\cal O}(k^3) \Bigg] .
\label{eq:tuvkerr4}
\eeal
The terms inside the parentheses look precisely like
an exponential completion of the expansion in \eqn{eq:cachazostrominger}.
Here it naturally appeared as a rewrite of the exponential structure
of the linearized Kerr energy-momentum tensor. We will see that this structure extends way beyond what is guaranteed by universality and it is a consequence of the `minimally coupled' nature of the Kerr solution. Note that the prefactor ${(p\cdot\varepsilon)^{2}}$
corresponds to the contribution
of the energy-momentum tensor of the linearized Schwarzschild solution \cite{Damour:2016gwp}.

Even though the fact that classical gravitational quantities can be reproduced from QFT computations has been known for a long time, the precise conceptual foundations of the matching are still lacking.\footnote{Very recent progress on relating
classical observables to quantum amplitudes has been made in~\cite{Kosower:2018adc}.} The goal of one of the authors in~\cite{Guevara:2017csg} was simply to show the agreement of the LS method 
with the previous computations of \cite{Holstein:2008sw,Holstein:2008sx,Vaidya:2014kza}. 
Moreover, in~\cite{Guevara:2017csg} the new massive spinor-helicity variables of Arkani-Hamed, Huang and Huang~\cite{Arkani-Hamed:2017jhn} were implemented to construct operators carrying spin multipoles. These operators were then matched, trough a change of basis, to those constructed
in~\cite{Holstein:2008sw,Holstein:2008sx,Vaidya:2014kza} in terms of polarization vectors
and Dirac spinors, enabling a systematic translation between the LS and the standard QFT amplitude in the $\hbar\to 0$ limit. It is only after computing the effective potential from this amplitude that one matches the post-Newtonian potential
of general relativity. 

The computation of the classical piece of the amplitude was made direct,
through the leading singularity, for arbitrary spin
and all orders in the center-of-mass energy $E$.
Both the tree-level and one-loop versions of this computation
correspond to a single order in the post-Minkowskian (PM) expansion
(see \eg recent discussion in~\cite{Damour:2016gwp,Bini:2017xzy,Vines:2017hyw,Damour:2017zjx,Bini:2018ywr,Bjerrum-Bohr:2018xdl,Cheung:2018wkq,Vines:2018gqi} and many more references therein), \ie at a fixed power of $G$.
However, the explicit match to the standard QFT amplitude
was only performed up to spin-1 and leading order in $E$
(which corresponds to the standard PN expansion).
Moreover, the computation of the PN effective potential
through the Born approximation suffers
some complications \cite{Holstein:2008sx,Neill:2013wsa}.
Such potential is not gauge-invariant, \ie not an observable,
and can undergo canonical and non-canonical transformations
that become cumbersome when spin is considered as part of the phase space.
Moreover, at one loop the Born approximation itself
requires the subtraction of tree-level pieces and suffers
from some (apparent) inconsistencies already at spin-1 \cite{Holstein:2008sw}.
For these reasons a more direct conversion from the LS
into a gravitational observable is evidently needed.
Very recently, a direct approach was proposed in the amplitudes setup
to evaluate the scattering angle of classical general relativity
\cite{Bjerrum-Bohr:2018xdl}, \ie the deflection angle of two
massive particles in the large-impact-parameter regime.
It was demonstrated that for scalar particles
the scattering angle computed by Westphal~\cite{Westpfahl:1985}
can be obtained via a simple 2D Fourier transform of the classical limit of the amplitude. 

Here we will show that the natural extension of the scattering angle,
for aligned spins as in~\cite{Bini:2017xzy,Vines:2017hyw,Bini:2018ywr,Vines:2018gqi},
can be computed with spinning particles directly from the LS.
The building blocks needed for this computation
are the three-point amplitude and the Compton amplitude
for massive spinning particles interacting with soft gravitons.
We will use the soft expansion with respect to the internal gravitons
to write the building blocks in an exponentiated form,
which fits naturally into the Fourier transform
leading to the first and second post-Minkowskian (1PM and 2PM) scattering angles
in a resummed form.

\subsection*{Summary of Results}

In \sec{sec:exponentiation3pt}
we show that the three-point scattering amplitude between
two massive particles of spin $s$ and one graviton is given by
\be
   {\cal M}_3^{(s)} (p_1,p_2,k^-)
    = \left(-\frac{\kappa}{2}\right)\times
      \frac{2(p\cdot\varepsilon)^2}{m^{2s}}\,\bra{2}^{2s}
      \exp\!\bigg(i\frac{k_\mu \varepsilon_\nu J^{\mu\nu}}
                        {p\cdot\varepsilon} \bigg)
      \ket{1}^{2s} , \qquad \quad
   p = \frac{p_1-p_2}{2} ,
\ee
where the exponential operator is generated by the angular momentum $J^{\mu\nu}$,
as appearing in the soft theorem \eqref{eq:cachazostrominger}.
This operator acts naturally on the product states
$|1\rangle^{2s}$ or $\ket{2}^{2s}$,
which are constructed from the new spinor-helicity variables
introduced by Arkani-Hamed, Huang and Huang~\cite{Arkani-Hamed:2017jhn}.
Denoting the operator by $\hat{\cal M}_3^{(s)}$ we write this as
\be
\hat{\cal M}_3^{(s)} = {\cal M}_3^{(0)} 
\exp\!\bigg(i\frac{k_\mu \varepsilon_\nu J^{\mu\nu}}{p\cdot\varepsilon}\bigg) ,
\label{M3op}
\ee
where ${\cal M}_3^{(0)}$ corresponds to
the amplitude for a massive scalar emitting a graviton.
In \sec{sec:exponentiation4pt} we extend this result
to the distinct-helicity Compton amplitude, showing that
\be
   {\cal M}_4^{(s)}(p_1,p_2,k_3^+,k_4^-)
    = \frac{1}{m^{2s}}\langle 2|^{2s} \hat{\cal M}_4^{(s)} |1\rangle^{2s} ,
   \qquad \quad
   \hat{\cal M}_4^{(s)}
    = {\cal M}_4^{(0)}
      \exp\!\bigg(i\frac{k_{4\mu} \varepsilon_{4\nu} J^{\mu\nu}}
                        {p\cdot\varepsilon_4}\bigg) ,
\label{comptonexponential}
\ee
up to corrections of fifth order in $J$ (appearing only for $s>2$).
In the operator form, $k_4$ and $\varepsilon_4$ can be replaced by
$k_3$ and $\varepsilon_3$,
which simply amounts to a change of basis.
The soft theorem~\eqref{eq:cachazostrominger} in this case is extrapolated in an exponential form, and corresponds to the simple statement of factorization of the Compton amplitudes into three-point amplitudes given by \eqn{M3op} and its plus-helicity version.

The formulas~\eqref{M3op} and~\eqref{comptonexponential} are the two building blocks needed to compute the scattering angle. In order to recover the classical observables we introduce and compute the {generalized expectation value} (GEV)
\be
   \braket{{\cal M}^{(s)}_n}
    = \frac{ \varepsilon_{2,\mu_1\ldots\mu_s}
             \hat{{\cal M}}^{\mu_1\ldots\mu_s,\nu_1\ldots\nu_s}_n
             \varepsilon_{1,\nu_1\ldots\nu_s} }
           { \varepsilon_2^{\mu_1\ldots\mu_s} \varepsilon_{1,\mu_1\ldots\mu_s} } 
    = \frac{ {\cal M}^{(s)}_n }
           { \varepsilon_2^{\mu_1\ldots\mu_s} \varepsilon_{1,\mu_1\ldots\mu_s} } .
\ee
Here we focus on integer-spin particles for simplicity,
therefore we use polarization tensors for spin $s$.
We first show that,
with $h_{\mu\nu} = 2\pi\delta(k^2)\varepsilon_\mu\varepsilon_\nu$,
\be
   h_{\mu\nu}(k) T^{\mu\nu}(-k)
    = \frac{1}{2} (2\pi)^2 \delta(k^2) \delta(p \cdot k)
      \lim_{s \rightarrow \infty} \braket{{\cal M}^{(s)}_3}
      \bigg|{\substack{~\\~\\p_1 =\,p-k/2~\\\:p_2 = -p-k/2}} ,
\label{eq:vertex2amp3}
\ee
where $T^{\mu\nu}$ on the LHS is the linearized stress-energy tensor of the Kerr black hole \eqref{eq:tuvkerr4}.
We then construct the aligned-spin scattering angle, for two-spinning-black-hole scattering,
as in~\cite{Kabat:1992tb,Akhoury:2013yua,Bjerrum-Bohr:2018xdl},
\be
   \theta = -\frac{E}{(2m_a m_b\gamma v)^2} \frac{\partial~}{\partial b}
      \int\!\frac{d^2\bs k}{(2\pi)^2}\,e^{i\bs k\cdot\bs b}
      \lim_{s_a,s_b \to \infty}  \braket{{\cal M}_4^{(s_a,s_b)}}
    + {\cal O}(G^3) 
\ee
(see \sec{sec:alignedspin} for definitions).
Here ${\cal M}_4^{(s_a,s_b)}$ corresponds to the four-point amplitude of figure \ref{fig:intro1}, with masses $m_a$ and $m_b$ and spin quantum numbers $s_a$ and $s_b$.  We compute this amplitude at both tree and one-loop levels using the LS proposed in~\cite{Guevara:2017csg}. The Fourier transform can be performed using the exponential forms~\eqref{M3op}-\eqref{comptonexponential}. 

We find the following expression for the aligned-spin scattering angle $\chi$ as a function of the masses $m_a$ and $m_b$, the rescaled spins (ring radii, intrinsic angular momenta per mass) $a_a$ and $a_b$, the relative velocity at infinity $v$, and the proper impact parameter $b$ (the impact parameter separating the zeroth-order/asymptotic worldlines defined by each black hole's Tulczyjew spin supplementary condition \cite{Tulczyjew:1959}):
\bse\label{finalfinalresult}
\be
   \theta = \frac{GE}{v^2}
      \bigg[ \frac{(1+v)^2}{b+a_a+a_b} + \frac{(1-v)^2}{b-a_a-a_b} \bigg]
    - \pi G^2 E\,\frac{\partial~}{\partial b}
      \bigg[ m_b f(a_a,a_b)+m_a f(a_b,a_a) \bigg] + {\cal O}(G^3) ,
\ee
where $E=\sqrt{m_a^2+m_b^2+2m_a m_b\gamma}$ with $\gamma=(1-v^2)^{-1/2}$, and
\be
\label{eq:fintro}
   f(\sigma,a) = \frac{1}{2a^2}
      \left({-b}+\frac{(\jmath+\varkappa-2a)^5}{4v\varkappa\big[(\jmath+\varkappa)^2-(2v a)^2\big]^{3/2}}
      \right) + {\cal O}(\sigma^5),
\ee
with
\be
   \jmath=vb+\sigma +a, \qquad \quad
   \varkappa=\sqrt{\jmath^2-4va(b+v\sigma)}.
\ee
\ese
This agrees with previous classical computations
to all orders in spin at tree level (at linear order in $G$) \cite{Bini:2017xzy,Vines:2017hyw}
and through linear order in spin at one loop (at order $G^2$) \cite{Bini:2018ywr},
as well as with the conjectural one-loop quadratic-in-spin expression presented in \cite{Vines:2018gqi}.
Moreover, \eqn{finalfinalresult} resums those contributions in a compact form, including higher orders in spin.
We have indicated that the expression~\eqref{eq:fintro} is valid up to quartic order in one of the spins (but to all orders in the other spin) according to the minimally coupled higher-spin amplitudes.

\section{Multipole expansion of three- and four-point amplitudes}
\label{sec:multipole}

\subsection{Massive spin-1 matter}
\label{sec:spin1matter}

We start our discussion of the multipole expansion
by dissecting the case of graviton emission by two massive vector fields.
The corresponding three-particle amplitude reads\footnote{We
omit the constant-coupling prefactors
$-(\kappa/2)^{n-2}$ in front of tree-level amplitudes,
we use $\kappa = \sqrt{32\pi G}$. Also note that we work
in the mostly-minus metric signature.}
\be
    {\cal M}_3(p_1,p_2,k)
     =-2(p\cdot\varepsilon)
       \big[ (p\cdot\varepsilon) (\varepsilon_1\!\cdot\varepsilon_2)
            - 2 k_\mu \varepsilon_\nu
              \varepsilon_1^{[\mu} \varepsilon_2^{\nu]}
       \big] , \qquad \quad
    p = \frac{1}{2} (p_1-p_2) ,
\label{eq:spin1amp3pt}
\ee
where $p$ is the average momentum of the spin-1 particle
before and after the graviton emission
and the polarization tensor of the graviton
$\varepsilon_{\mu\nu}=\varepsilon_\mu\varepsilon_\nu$ 
(with momentum $k=-p_1-p_2$)
is split into two massless polarization vectors.
The derivation of \eqn{eq:spin1amp3pt} from the Proca action is detailed in \app{app:spin1amp3pt},
which also motivates that the term involving
$\varepsilon_1^{[\mu} \varepsilon_2^{\nu]}$
can be thought of as an angular-momentum contribution to the scattering.
In other words, we are tempted to interpret
the combination $\varepsilon_1^{[\mu} \varepsilon_2^{\nu]}$
as being (proportional to) the classical spin tensor.

However, we now face our first challenge:
as explained in~\cite{Holstein:2008sw,Holstein:2008sx,Vaidya:2014kza},
the spin-1 amplitude contains up to quadrupole interactions,
\ie quadratic in spin,
whereas only the linear piece is apparent in \eqn{eq:spin1amp3pt}.
To rewrite this contribution in terms of multipoles,
we can use a redefined spin tensor
\be
   S^{\mu\nu}
    =-\frac{i}{\varepsilon_1\!\cdot \varepsilon_2} \bigg\{
      2\varepsilon_1^{[\mu} \varepsilon_2^{\nu]}
    - \frac{1}{m^2} p^{[\mu}
      \big( (k\cdot\varepsilon_2) \varepsilon_1
          + (k\cdot\varepsilon_1) \varepsilon_2 \big)^{\nu]} \bigg\} .
\label{eq:spintdef}
\ee
It is introduced in \app{app:spin1tensor}
via a two-particle expectation value/matrix element,
which we call the generalized expectation value (GEV)
\begin{equation}
S^{\mu\nu}
=\frac{\varepsilon_{2\sigma} \hat{\Sigma}^{\mu\nu,\sigma}_{~~~~\;\tau}
       \varepsilon_1^\tau} {\varepsilon_{2\sigma}\varepsilon_1^\sigma} .
\label{gevmotiv}
\end{equation}
Here $\hat{\Sigma}^{\mu\nu}$ is constructed as an angular-momentum operator
shifted in such a way that its GEV satisfies the Fokker-Tulczyjew covariant
spin supplementary condition (SSC)~\cite{Fokker:1929,Tulczyjew:1959}
\be
    p_\mu S^{\mu\nu} = 0 .
\label{eq:ssc}
\ee
In this paper we find this condition to be crucial for the matching
to the rotating-black-hole computation of \cite{Vines:2017hyw},
as the classical spin tensor $S^{\mu\nu}$~\eqref{eq:spinvector2spintensor}
satisfies the above SSC by definition.
The purpose of this SSC is to constrain the mass-dipole components~$S^{0i}$
of the spin tensor of an object to vanish in its rest frame.
In a classical setting it puts the reference point
for the intrinsic spin of a spatially extended object
at its rest-frame center of mass.

Inserting this spin tensor in \eqn{eq:spin1amp3pt2},
we rewrite the above amplitude as
\beal
    {\cal M}_3(p_1,p_2,k) &
      =-m^2 x^2 (\varepsilon_1\!\cdot\varepsilon_2)
        \bigg[ 1 - \frac{i\sqrt{2}}{m x} k_\mu \varepsilon_\nu S^{\mu\nu}
             + \frac{ (k\cdot\varepsilon_1) (k\cdot\varepsilon_2) }
                    { m^2 (\varepsilon_1\!\cdot\varepsilon_2) }
        \bigg] ,
\label{eq:spin1amp3pt2}
\eeal
where for further convenience we also expressed
the scalar products $p\cdot\varepsilon$ using
a helicity variable~$x$ first introduced in~\cite{ArkaniHamed:2008gz}
\be
   x = \sqrt{2}\,\frac{p\cdot\varepsilon}{m}
\label{eq:xvariable}
\ee
(at higher points it becomes gauge-dependent but can still be used as a shorthand).
Now, in the GEV of the amplitude,
\be
\braket{\mathcal{M}_3}
 = \frac{ \varepsilon_{2\sigma} \mathcal{M}_3^{\sigma \tau} \varepsilon_{1,\tau} }
        { \varepsilon_{2\sigma} \varepsilon_1^\sigma }
 = -m^2 x^2 
   \bigg[ 1 - i\frac{k_\mu \varepsilon_\nu S^{\mu\nu}}{p\cdot \varepsilon} 
            + \frac{ (k\cdot\varepsilon_1) (k\cdot\varepsilon_2) }
                   { m^2 (\varepsilon_1\!\cdot\varepsilon_2) } \bigg],
\label{gevmm}
\ee
we recognize the dipole coupling of \eqn{eq:tuvkerr4} as
the term linear in both $k$ and $S$.
Indeed, particles with spin couple naturally
to the field-strength tensor of the graviton
$F_{\mu\nu}=2k_{[\mu}\varepsilon_{\nu]}$, analogously
to the magnetic dipole moment $F_{\mu\nu} S^{\mu\nu}$.\footnote{We thank
Yu-tin Huang for emphasizing to us the analogy
to the electromagnetic Zeeman coupling, see e.g. \cite{Holstein:2006pq,Goldberger:2017ogt}. Indeed, in a non-covariant form, this was already related to the soft expansion long ago \cite{osti_4073049}.}
Following the non-relativistic limit, the third term was identified
in~\cite{Holstein:2008sw,Holstein:2008sx,Vaidya:2014kza,Guevara:2017csg}
to be the quadrupole interaction
$\propto \big(F_{\mu\nu} S^{\mu \nu} \big)^2$ for spin-1.
It may seem a priori puzzling that we wish to regard the interaction
$(k \cdot \varepsilon_1) (k \cdot \varepsilon_2)$
as the square of $F_{\mu\nu} S^{\mu\nu}$.
This is because the statement is true at the levels of spin operators,
but not at the level of (generalized) expectation values, \ie
$\langle F_{\mu\nu} \hat{\Sigma}^{\mu \nu}\rangle^2 \neq 
 \langle \big(F_{\mu\nu} \hat{\Sigma}^{\mu \nu}\big)^2 \rangle$.
In order to expose the exponential structure described in the introduction
and construct such spin operators at any order,
we are going to recast the multipole expansion
in terms of spinor-helicity variables.

\subsubsection{Spinor-helicity recap}

This subsection can be skipped if the reader is familiar
with the massive spinor-helicity formalism
of Arkani-Hamed, Huang and Huang~\cite{Arkani-Hamed:2017jhn},\footnote{The spinor-helicity conventions used in the present paper are detailed in the latest arXiv version of \cite{Ochirov:2018uyq}.}
which is well suited to describe scattering amplitudes
for massive particles with spin.
Much like its massless counterpart,
this formalism allows to construct all of the scattering kinematics
from basic ${\rm SL}(2,\mathbb{C})$ spinors that transform covariantly
with respect to the little group of the associated particle.
The massive little group is ${\rm SU}(2)$,
so the Pauli-matrix map from two-spinors to momenta
\be
    p_{\alpha\dot{\beta}} = p_{\mu} \sigma^\mu_{\alpha\dot{\beta}}
    = \epsilon_{ab} \ket{p^a}_\alpha [p^b|_{\dot{\beta}}
    = \ket{p^a}_\alpha  [p_a|_{\dot{\beta}}
    = \la_{\alpha}^{~a} \lb_{\dot{\beta}a} ,
\label{eq:massivespinors}
\ee
involves a contraction of the ${\rm SU}(2)$ indices $a,b,\ldots=1,2$
(not to be confused with the spinorial ${\rm SL}(2,\mathbb{C})$ indices
$\alpha,\beta,\ldots=1,2$ and $\dot{\alpha},\dot{\beta},\ldots=1,2$).
This is in contrast to the massless case,
where the little group is ${\rm U}(1)$, so its index is naturally hidden
inside the complex nature of massless two-spinors
\be
    k_{\alpha\dot{\beta}} = k_\mu \sigma^\mu_{\alpha\dot{\beta}}
    = \ket{k}_\alpha  [k|_{\dot{\beta}}
    = \la_\alpha  \lb_{\dot{\beta}} .
\ee
Now just as $\la_\alpha$ and $\lb_{\dot{\beta}}$
are convenient to built massless polarization vectors~\eqref{eq:xvariables}, we can use the massive spinors
$\la_{\alpha}^{~a}$ and $\lb_{\dot{\beta}}^{~b}$
to construct spin-$S$ external wavefunctions.
For instance, massive polarization vectors are explicitly
\be
   \varepsilon_{p\:\!\mu}^{ab}
    = \frac{\bra{p^{(a}}\sigma_\mu|p^{b)}]}{\sqrt{2}m}
   \qquad \Rightarrow \qquad
   \left\{
   \begin{aligned}
   p\cdot\varepsilon_{p}^{ab} & = 0 ,\\
   \varepsilon_{p\:\!\mu}^{ab} \varepsilon_{p\:\!\nu ab} &
    = \eta_{\mu\nu} - \frac{p_\mu p_\nu}{m^2} \\
   \varepsilon_{p\:\!11}\!\cdot\!\varepsilon_{p}^{11} &
    = \varepsilon_{p\:\!22}\!\cdot\!\varepsilon_{p}^{22}
    = 2\:\!\varepsilon_{p\:\!12}\!\cdot\!\varepsilon_{p}^{12} = 1 ,
   \end{aligned}
   \right.
\label{eq:polvectorsmassive}
\ee
where the symmetrized little-group indices $(ab)$
represent the physical spin-projection numbers $1,0,-1$ with respect
to a spin quantization axis, as chosen by the massive spinor basis.
Note that the vector indices,
as well as their dotted and undotted spinorial counterparts,
must always be contracted and do not represent a physical quantum number.

Let us also point out here that the massless polarization vectors
and hence the associated helicity variable~\eqref{eq:xvariable}
can be written in terms of massless spinors as
\be
   \varepsilon_+^\mu
    = \frac{\bra{r}\sigma^{\mu}|k]}{\sqrt{2}\braket{r\;\!k}} ,
   \qquad \quad
   \varepsilon_-^\mu
    =-\frac{[r|\bar{\sigma}^{\mu}\ket{k}}{\sqrt{2}[r\;\!k]}
   \qquad \Rightarrow \qquad
    x_+ = \frac{\bra{r}p|k]}{m\braket{r\;\!k}} , \qquad
    x_- =-\frac{[r|p\ket{k}}{m[r\;\!k]} = -\frac{1}{x_+} ,
\label{eq:xvariables}
\ee
where $x$ is independent of the reference momentum $r$
on the three-point on-shell kinematics.

\subsubsection{Spin-1 amplitude in spinor-helicity variables}
\label{sec:spin1amp}

We can now obtain concrete spinor-helicity expressions
for the amplitude~\eqref{eq:spin1amp3pt}.
Choosing the polarization of the graviton to be negative, we have
\begin{subequations}
\begin{align}
\label{eq:vectors2spinors0}
    \varepsilon_1^{a_1 a_2}\!\cdot\varepsilon_2^{b_1 b_2} &
    = -\frac{1}{m^2} \braket{1^{(a_1} 2^{(b_1}}
      \bigg[ \braket{1^{a_2)} 2^{b_2)}}
           - \frac{1}{m x} \braket{1^{a_2)} k} \braket{k\;\!2^{b_2)}}
      \bigg] , \\ \!
\label{eq:vectors2spinors1}
    \big[(\varepsilon_1\!\cdot\varepsilon_2)
    k_\mu \varepsilon_\nu^- S^{\mu\nu}\big]^{a_1 a_2 b_1 b_2} &
    = \frac{i}{\sqrt{2}m^2} \braket{1^{(a_1}k}
      \bigg[ \braket{1^{a_2)} 2^{(b_1}}
           - \frac{1}{2m x} \braket{1^{a_2)} k} \braket{k\;\!2^{(b_1}}
      \bigg] \braket{k\;\!2^{b_2)}} , \\
\label{eq:vectors2spinors2}
    (k\cdot\varepsilon_1^{a_1 a_2}) (k\cdot\varepsilon_2^{b_1 b_2}) &
    =-\frac{1}{2m^2 x^2} \braket{1^{(a_1}k} \braket{1^{a_2)}k}
      \braket{k\;\!2^{(b_1}} \braket{k\;\!2^{b_2)}} ,
\end{align} \label{eq:vectors2spinors}%
\end{subequations}
where we have reduced all $[1^a|$ and $|2^b]$
to the chiral spinor basis of $\bra{1^a}$ and $\ket{2^b}$
using the following identities for the three-point kinematics,\footnote{The
transition between the chiral spinors $\ket{p^a}$
and the antichiral ones $|p^a]$
is always possible~\cite{Arkani-Hamed:2017jhn} via the Dirac equations
$p^{\dot{\alpha}\beta} \ket{p^a}_{\;\!\!\beta} = m |p^a]^{\dot{\alpha}}$ and
$p_{\alpha\dot{\beta}} |p^a]^{\dot{\beta}} = m \ket{p^a}_{\;\!\!\alpha}$.
}
\be
    [1^a k] = x^{-1} \braket{1^a k} , \qquad \quad
    [2^b k] =-x^{-1} \braket{2^b k} , \qquad \quad
    [1^a 2^b] = \braket{1^a 2^b}
    - \frac{1}{m x} \braket{1^a k} \braket{k\;\!2^b} .
\label{eq:squaretoanglethree-point}
\ee
We also use $x$ for $x_-$ henceforth,
\ie it carries helicity $-1$ unless stated otherwise.
From \eqn{eq:vectors2spinors} we can see that
going to the chiral spinor basis has both an advantage and a disadvantage.
On the one hand, the multipole expansion becomes transparent
in the sense that the spin order of a term
is identified by the leading power of $\ket{k}\bra{k}$.
On the other hand, the exponential structure of the vector basis
is spoiled by a shift by higher multipole terms.
However, this is just an artifact of the chiral basis,
and we should see that the answer obtained
from the generalized expectation value is the same.

The main advantage of the spinor-helicity variables
for what we wish to achieve in this paper is that
now we can switch to spinor tensors $\bra{1^{(a_1}}\otimes\bra{1^{a_2)}}$
and $\ket{2^{(b_1}}\otimes\ket{2^{b_2)}}$,
as representations of the massive-particle states 1 and 2.
Introducing the symbol $\odot$ for the symmetrized tensor product,
we can rewrite \eqn{eq:vectors2spinors0} as
\be
    \varepsilon_1\cdot\varepsilon_2
    = -\frac{1}{m^2} \bra{1}^{\odot 2}
       \bigg[ \mathbb{I} \odot \mathbb{I}
            - \frac{1}{m x} \mathbb{I} \odot \ket{k} \bra{k}
       \bigg] \ket{2}^{\odot 2}
    = -\frac{1}{m^2}
       \bigg[ \braket{12}^{\odot 2}
            - \frac{1}{m x} \braket{12} \odot \braket{1k} \braket{k2}
       \bigg] .
\ee
Here the operators have their lower indices symmetrized, \ie
$(A \odot B)_{\alpha_1\beta_1}^{\alpha_2\beta_2}
=A_{(\alpha_1}^{\beta_1} B_{\alpha_2)}^{\beta_2}$,
and the notation assumes that the reader keeps in mind the spins
associated with each momentum.
Combining all the terms in \eqn{eq:vectors2spinors}
into the amplitude, we obtain
\beal
{\cal M}_3(p_1,p_2,k^-) & = x^2
    \bigg[ \braket{12}^{\odot 2}
         - \frac{2}{m x} \braket{12}\!\odot\!\braket{1k}\braket{k2}
         + \frac{1}{m^2 x^2}
           \braket{1k}^{\odot 2} \braket{k2}^{\odot 2}
    \bigg] .
\label{eq:spin1amp3pt3}
\eeal

Now in the multipole expansion
of the Kerr stress-energy tensor~\eqref{eq:tuvkerr4},
the quadrupole operator is of the simple form
$\left(k_\mu \varepsilon_\nu S^{\mu\nu}\right)^2$,
whereas in our amplitude~\eqref{eq:spin1amp3pt2}
it has the form $(k\cdot\varepsilon_1)(k\cdot\varepsilon_2) \propto \braket{1k}^{\odot 2} \braket{k2}^{\odot 2}$.
One then could wonder if in some sense the latter
is the square of $(k_\mu \varepsilon_\nu S^{\mu\nu})$.
We now show that this is precisely the case if the angular momentum
is realized as a differential operator.

In \app{app:angularmomentum} we construct the differential form
of the angular-momentum operator in momentum space starting from its
definition
\be
J^{\mu\nu} = i p^\mu \frac{\partial~}{\partial p_\nu}
           - i p^\nu \frac{\partial~}{\partial p_\mu} + \text{intrinsic} ,
\ee
which involves the standard orbital piece
and the ``intrinsic'' contribution dependent on spin.
This operator admits a much simpler realization in terms of spinor variables,
similar to the one derived in~\cite{Witten:2003nn} for the massless case.
For a massive particle of momentum
$p_{\alpha\dot{\beta}} = \la_{p\:\!\alpha}^{~\;a} \lb_{p\:\!\dot{\beta}a}$
we find that the differential operator
for the total angular momentum is given by
\be
J_{\alpha\dot{\alpha},\beta\dot{\beta}}
= 2i\bigg[ \la_{p(\alpha}^{~~a} \frac{\partial~}{\partial \la_p^{\beta)a}}
           \epsilon_{\dot{\alpha}\dot{\beta}}
         + \epsilon_{\alpha\beta} \lb_{p(\dot{\alpha}}^{~~a}
           \frac{\partial~}{\partial \lb_p^{\dot{\beta})a}} \bigg] .
\label{eq:totalspinormassive}
\ee

We can now act with the operator $k_\mu \varepsilon_\nu J^{\mu\nu}$
on the product state
$\ket{p^a}^{\odot 2}=\ket{p^{a_1}}\otimes\ket{p^{a_2}}$.
For the negative helicity of the graviton, we have
\be
k_\mu \varepsilon_\nu^- J^{\mu\nu}
= \frac{1}{4\sqrt{2}} \la^\alpha \la^\beta
  \epsilon^{\dot{\alpha}\dot{\beta}} J_{\alpha\dot{\alpha},\beta\dot{\beta}}
= -\frac{i}{\sqrt{2}} \braket{k\;\!p^a}
  \braket{k\;\!\frac{\partial~}{\partial \la_p^a}} , \qquad \qquad
  \braket{k\;\!\frac{\partial~}{\partial \la_p^b}} \ket{p^a}
= \ket{k}\delta^a_b .
\ee
Applying the spinor differential operator above, we find\footnote{More explicitly,
we have
\begin{align*}
i\sqrt{2} (k_\mu \varepsilon_\nu^- J^{\mu\nu}) \ket{p^a}^{\odot 2} &
 = \braket{k\;\!p^b} \bigg\{ 
  \bigg[ \braket{k\;\!\frac{\partial~}{\partial \la_p^b}} \ket{p^{a_1}}
  \bigg]\!\otimes\!\ket{p^{a_2}}
+ \ket{p^{a_1}}\!\otimes\!
  \bigg[ \braket{k\;\!\frac{\partial~}{\partial \la_p^b}} \ket{p^{a_2}}
  \bigg] \bigg\} \\ &
= \ket{k} \braket{k\;\!p^{a_1}}\!\otimes\!\ket{p^{a_2}}
+ \ket{p^{a_1}}\!\otimes\!\ket{k}\braket{k\;\!p^{a_2}}
= 2 \ket{k} \braket{k\;\!p^a}\!\odot\!\ket{p^a} ,
\end{align*}
with similar manipulations for higher powers.
}
\begin{subequations} \begin{align}
\bigg(\frac{i k_\mu \varepsilon_\nu^- J^{\mu\nu}}{p\cdot\varepsilon^-}
\bigg) \ket{p}^2 &
 = \frac{2}{m x} \ket{k} \braket{k\;\!p} \ket{p} ,
\label{eq:raising1} \\
\bigg(\frac{i k_\mu \varepsilon_\nu^- J^{\mu\nu}}{p\cdot\varepsilon^-}
\bigg)^{\!2} \ket{p}^2 &
= \frac{2}{m^2 x^2} \ket{k}^2 \braket{k\;\!p}^{2} ,
\label{eq:raising2} \\
\bigg(\frac{i k_\mu \varepsilon_\nu^- J^{\mu\nu}}{p\cdot\varepsilon^-}
\bigg)^{\!j} \ket{p}^2 & = 0 ,
  \qquad \qquad \qquad \qquad \qquad j \geq 3 .
\label{eq:raising3}
\end{align} \label{eq:raising}%
\end{subequations}
Although it is the differential operator that realizes the soft theorem,
its algebraic form is easy to obtain on three-particle kinematics.
Indeed, if we take a tensor-product version
$-(\sigma^{\mu\nu}\otimes\mathbb{I} + \mathbb{I}\otimes\sigma^{\mu\nu})$
of the standard ${\rm SL}(2,\mathbb{C})$ chiral generator
$\sigma^{\mu\nu}=i\sigma^{[\mu}\bar{\sigma}^{\nu]}/2$
and use it as an algebraic realization of $J^{\mu\nu}$,
it is direct to check that
it acts in the same way as the differential operator above:
\be
\frac{i k_\mu \varepsilon_\nu^- J^{\mu\nu}}{p\cdot\varepsilon^-}
= \frac{\ket{k} \bra{k}}{m x} \otimes \mathbb{I}
+ \mathbb{I} \otimes \frac{\ket{k} \bra{k}}{m x} .
\label{eq:linspi}
\ee

These identities allow us to reinterpret
the last two terms in the amplitude formula~\eqref{eq:spin1amp3pt3}
as the non-zero powers of this dipole operator
acting on the state $\ket{1}^2$:
\be
  -\frac{2}{mx} \braket{12}\braket{1k}\braket{k2} = \bra{2}^2
   \bigg( \frac{ik_\mu \varepsilon_\nu^- J_1^{\mu\nu}}
               {p_1\!\cdot\varepsilon^-} \bigg) \ket{1}^2 , \qquad \quad
   \frac{1}{m^2 x^2} \braket{1k}^2 \braket{k2}^2 = \frac{1}{2} \bra{2}^2
   \bigg( \frac{ik_\mu \varepsilon_\nu^- J_1^{\mu\nu}}{p_1\cdot\varepsilon^-}
   \bigg)^{\!2} \ket{1}^2 ,
\label{eq:spin1multipoles}
\ee
and rewrite the amplitude as
\be
{\cal M}_3(p_1,p_2,k^-)
  = x^2 \bra{2}^2 \bigg\{ 1
  + i\bigg(\frac{k_\mu \varepsilon_\nu^- J_1^{\mu\nu}}
               {p_1\cdot\varepsilon^-}\bigg)
  - \frac{1}{2}
    \bigg(\frac{k_\mu \varepsilon_\nu^- J_1^{\mu\nu}}
               {p_1\cdot\varepsilon^-}\bigg)^{\!2}
    \bigg\} \ket{1}^2 .
\ee
It is now clear that these terms
\begin{itemize}
\item match the differential operators
of the soft expansion~\eqref{eq:cachazostrominger};
\item correspond to the scalar, spin dipole and quadrupole interactions
in the expansion of the Kerr energy momentum tensor~\eqref{eq:tuvkerr4}
and its spin-1 amplitude representation~\eqref{gevmm}.
Note that the sign flip in the dipole term comes from the sign difference
between the algebraic and differential Lorentz generators,
as pointed out in the beginning of \app{app:angularmomentum}.
\end{itemize}
In this way, we interpret the three terms
in the amplitude~\eqref{eq:spin1amp3pt3}
as the multipole contributions
with respect to the chiral spinor basis,
despite the fact that they do not equal
the multipoles in \eqn{eq:spin1amp3pt2} individually.
Furthermore, as the operator $(k_\mu \varepsilon_\nu^-J^{\mu\nu})^j$
annihilates the spin-1 state for $j \geq 3$,
the three terms can be obtained from an exponential
\be
{\cal M}_3(p_1,p_2,k^-) = x^2 \bra{2}^2
    \exp\!{\bigg(i\frac{k_\mu \varepsilon^-_\nu J^{\mu\nu}}
                       {p\cdot\varepsilon^-}} \bigg) \ket{1}^2 .
\label{eq:m3m}
\ee

It can be checked explicitly that acting with the operator on the
state $\bra{2}^2$ yields the same result,
\ie in this sense the operator
$k_\mu \varepsilon_\nu J^{\mu\nu}/(p\cdot\varepsilon)$
is self-adjoint.\footnote{The division by $p\cdot\varepsilon$ implicitly
relies on the fact that the action of $k_\mu \varepsilon_\nu J^{\mu\nu}$
on the helicity variable $x$ vanishes.
Note also that $k_\mu \varepsilon_\nu J^{\mu\nu}/(p\cdot\varepsilon)$
should become $k_\mu \varepsilon_\nu J_2^{\mu\nu}/(p_2\cdot\varepsilon)$
when acting on $\ket{2}^2$.
}
On the other hand,
choosing the other helicity of the graviton will yield the parity
conjugated version of \eqn{eq:m3m}:
\be
{\cal M}_3(p_1,p_2,k^+) = \frac{1}{x^2} [2|^2
    \exp\!\bigg(i\frac{k_\mu \varepsilon^+_\nu J^{\mu\nu}}
                      {p\cdot\varepsilon^+} \bigg) |1]^2 .
\label{eq:m3p}
\ee

In the next section we extend this procedure to arbitrary spin.
Let us point out that the explicit amplitude
can be brought into a compact form by changing the spinor basis.
In fact, the three-point identities~\eqref{eq:squaretoanglethree-point} imply that
the amplitude formula~\eqref{eq:spin1amp3pt3} collapses into
\be
{\cal M}_3(p_1,p_2,k^-) = [12]^2 x^2 .
\label{mincoups1}
\ee
However, let us stress that this form completely hides the spin structure that was already explicit in the vector form~\eqref{eq:spin1amp3pt2}.
The purpose of the insertion of the differential operators is precisely to extract the spin-dependent pieces from the minimal coupling~\eqref{mincoups1},
which will then be matched to the Kerr black hole.

\subsection{Exponential form of three-particle amplitude}
\label{sec:exponentiation3pt}

In this section we generalize the previous discussion to arbitrary spin~$s$.
Concentrating our attention on integer spin
allows us to ignore factors of $(-1)^{2s}$.
The starting point in this case is
the three-point amplitudes for massive matter minimally coupled to gravity
in the little-group sense~\cite{Arkani-Hamed:2017jhn}:
\be
{\cal M}_3^{(s)}(p_1,p_2,k^+)
= \frac{\braket{12}^{2s} x^{-2}}{m^{2s-2}} , \qquad \quad
{\cal M}_3^{(s)}(p_1,p_2,k^-)
= \frac{[12]^{2s} x^2}{m^{2s-2}} .
\ee
As explained in the previous section, in such a compact form
all the dependence on the spin tensor is completely hidden.
In order to restore it,
we need to write the minus-helicity amplitude in the chiral basis
\beal
{\cal M}_3^{(s)}(p_1,p_2,k^-)
= \frac{x^2}{m^{2s-2}}
\bigg( \braket{21} + \frac{\braket{2k}\braket{k1}}{m x}
\bigg)^{\!\odot 2s}\!
= \frac{x^2}{m^{2s-2}} \bra{2}^{2s}
\Bigg[ \sum_{j=0}^{2s} \binom{2s}{j}
       \bigg( \frac{\ket{k}\bra{k}}{m x}\bigg)^j \Bigg] \ket{1}^{2s} ,
\label{eq:bin}
\eeal
where we have taken advantage of the symmetrized tensor product $\odot$
that enables us to perform the binomial expansion
(we have suppressed the identity factors in the tensor product).
Even though this already corresponds to an expansion
in the ``spin operator'' of \cite{Guevara:2017csg},
here we recast this into exponential form
by inserting the differential angular momentum operator
\be
i\frac{k_\mu \varepsilon^-_\nu J^{\mu\nu}}{p\cdot\varepsilon^-}
= \frac{1}{m x} \braket{k\;\!p}
  \braket{k\;\!\frac{\partial~}{\partial \la_p}} ,
\qquad \quad
\braket{k\;\!p} \braket{k\;\!\frac{\partial~}{\partial \la_p}} \ket{p}
= \ket{k} \braket{k\;\!p} .
\label{eq:diff2raising}
\ee
Indeed, it is easy to generalize the formulae~\eqref{eq:raising}
to product states of spin-$s$, namely
\be
\bigg(i\frac{k_\mu \varepsilon^-_\nu J^{\mu\nu}}
            {p\cdot\varepsilon^-}\bigg)^{\!j} \ket{p}^{2s} =
\left\{
\begin{array}{ll}
\dfrac{(2s)!}{(2s-j)!} \ket{p}^{2s-j}
\bigg( \dfrac{\ket{k} \braket{k\;\!p}}{mx}
\bigg)^{\!j} &, \quad j \leq 2s , \\
0 &, \quad j > 2s .
\end{array}
\right.
\label{eq:raisinggeneral}
\ee
In other words,
in general the operator~\eqref{eq:diff2raising} is nilpotent
of order $2s$.\footnote{Interestingly,
due to its property~\eqref{eq:raisinggeneral}
the spinorial differential operator~\eqref{eq:diff2raising}
can be regarded as a ladder operator for a spin-$s$ representation.}
Of course, this also admits an algebraic realization,
which is extends the formula~\eqref{eq:linspi}.
From this we can derive the formal relations\footnote{For $j=1$,
\eqn{eq:raisinggeneral2} corresponds to the operator $k \cdot S$
used in~\cite{Guevara:2017csg} to perform
the matching with the standard QFT amplitude.
We note, however, that the classical quantity
${k_\mu \varepsilon_\nu S^{\mu\nu}}/{(p\cdot\varepsilon)}$
matches the quantity $k \cdot S$ used in~\cite{Guevara:2017csg}
only when the spin tensor satisfies the SSC~\eqref{eq:ssc},
as can be seen by squaring both terms.}
\be
\bigg(i\frac{k_\mu \varepsilon^-_\nu 
             J^{\mu\nu}}{p\cdot\varepsilon^-}\bigg)^{\!\odot j}  =
\left\{
\begin{array}{ll}
\dfrac{(2s)!}{(2s-j)!}
\bigg( \dfrac{\ket{k} \bra{k}}{mx}
\bigg)^{\!\otimes j}\!\odot \mathbb{I}^{\otimes 2s-j} &, \quad j \leq 2s , \\
0 &, \quad j > 2s .
\end{array}
\right.
\label{eq:raisinggeneral2}
\ee
Therefore, we can rewrite \eqn{eq:bin} as an exponential
\be
\bra{2}^{2s}
\Bigg[ \sum_{j=0}^{2s} \binom{2s}{j}
       \bigg( \frac{\ket{k}\bra{k}}{m x}\bigg)^j \Bigg] \ket{1}^{2s}\!
  = \bra{2}^{2s}\!\sum_{j=0}^{\infty}\frac{1}{j!}
    \bigg( i\frac{k_\mu \varepsilon^-_\nu J^{\mu\nu}}
                 {p\cdot\varepsilon^-}\bigg)^{\!j} \ket{1}^{2s}\!
  = \bra{2}^{2s}\!
    \exp\!\bigg(i\frac{k_\mu \varepsilon^-_\nu J^{\mu\nu}}
                      {p\cdot\varepsilon^-}\bigg) \ket{1}^{2s} ,
\label{eq:bin2exp}
\ee
where we note that the exponential expansion,
albeit valid to all orders, becomes trivial at order~$2s$.
It can be read from \eqn{eq:bin} that
the spin operator $\ket{k}\bra{k}$ of \cite{Guevara:2017csg}
corresponds precisely to $k_\mu \varepsilon^-_\nu J^{\mu \nu}$.
Moreover, in the formal limit~$s\rightarrow \infty$
the exponential can be realized as a linear operator that does not truncate!
However, let us stress that even for finite spins
the exponential operator in
\be
\hat{{\cal M}}_3^{(s)}(p_1,p_2,k^-) = {\cal M}_3^{(0)}
   \exp\!\bigg(i\frac{k_\mu \varepsilon^-_\nu J^{\mu\nu}}
                     {p\cdot\varepsilon^-}\bigg) ,
\qquad \quad
{\cal M}_3^{(s)} = \frac{1}{m^{2s}}
    \bra{2}^{2s}\hat{{\cal M}}_3^{(s)}\ket{1}^{2s}
\label{eq:exp3m}
\ee
is still present and can be mapped to classical observables
such as the scattering angle.
This framework will be particularly useful at order $G^2$,
since the arbitrary spin version (and hence the $s\rightarrow\infty$ limit) of the Compton amplitude is not yet known.

Analogously, it can be shown that the transition to the positive helicity
amounts to exchanging angle brackets with square brackets:
\be
\hat{{\cal M}}_3^{(s)}(p_1,p_2,k^+) = {\cal M}_3^{(0)}
   \exp\!\bigg(i\frac{k_\mu \varepsilon^+_\nu J^{\mu\nu}}
                     {p\cdot\varepsilon^+}\bigg) ,
\qquad \quad
{\cal M}_3^{(s)} = \frac{1}{m^{2s}}
    [2|^{2s}\hat{{\cal M}}_3^{(s)}|1]^{2s} .
\label{eq:exp3p}
\ee

The forms~\eqref{eq:exp3m} and~\eqref{eq:exp3p} make explicit the fact
that the higher-spin amplitude is non-local~\cite{Arkani-Hamed:2017jhn}.
However, despite the appearance of the factor $p\cdot\varepsilon$
in the denominator, the exponential factor is gauge-invariant
due to the three-particle kinematics.
We further recognize in the argument of the exponential the same structure
as the one appearing in the Cachazo-Strominger soft theorem.
In fact, as will be made explicit in the next section,
the extended soft factor of Cachazo and Strominger
is just an instance of a three-point amplitude of higher-spin particles.
The poles present in the extended soft factor~\eqref{eq:cachazostrominger}
simply arise when gluing these three-point amplitudes.

The formula \eqref{eq:exp3m} is our first main result.
Note that this holds for the full three-point amplitude
with no classical limit whatsoever.
This formula matches precisely
the Kerr energy-momentum tensor \eqref{eq:tuvkerr4},
with ${\cal M}_3^{(0)} = m^2 x^2$ corresponding to the scalar piece
(the Schwarsczhild case).
In \sec{sec:HCL} we will use this compact form to compute
the scattering angle of two Kerr black holes at linear order in $G$.

\subsection{Exponential form of gravitational Compton amplitude}
\label{sec:exponentiation4pt}

The task of this section is to extend the construction presented in
the previous one to the Compton amplitude,
without the support of three-particle kinematics.\footnote{Historically, the Compton amplitude was the prototype in the discovery of subleading soft theorems \cite{Low:1954kd,Gross:1968in,Jackiw:1968zza}. The construction provided in section \ref{sec:factorization} is in a sense reminiscent of Low's original derivation of the subleading factor in QED \cite{Low:1954kd}.}
In particular, we will show that
for the cases of interest the following holds
\be
\hat{{\cal M}}_4^{(s)}(p_1,p_2,k_3^+,k_4^-) = {\cal M}_4^{(0)}
\exp\!\bigg(i\frac{k_\mu \varepsilon_\nu J^{\mu\nu}}
                  {p\cdot\varepsilon}\bigg) .
\label{eq:compexp}
\ee
Here the linear and angular momentum~$p$ and $J^{\mu\nu}$
in the exponential operator may act either on massive state~$1$ or~$2$.
Moreover, the momentum~$k$ and the polarization vector~$\varepsilon$
can be associated to either of the two gravitons.
Explicitly, we have
\be
[2|^{2s}
\exp\!\bigg(i\frac{k_{3\mu}\varepsilon_{3\nu}^+J^{\mu\nu}}
                  {p\cdot\varepsilon_3^+}\bigg) |1]^{2s} =
\bra{2}^{2s}
\exp\!\bigg(i\frac{k_{4\mu}\varepsilon_{4\nu}^-J^{\mu\nu}}
                  {p\cdot\varepsilon_4^-}\bigg) \ket{1}^{2s} .
\label{eq:conj}
\ee

The importance of this amplitude (as opposed to the same-helicity
case) is that it controls the classical contribution at order $G^{2}$,
as was shown directly in~\cite{Guevara:2017csg,Bjerrum-Bohr:2018xdl}.
in~\cite{Guevara:2017csg}
the classical piece was argued to lead to the correct 2PN potential
after a Fourier transform. In the new approach of \cite{Bjerrum-Bohr:2018xdl} the
classical contribution in the spinless case was identified by computing
the scattering angle. In \sec{sec:HCL} we will use the Compton amplitude
as an input for computing the scattering angle with spin up to order
$S^{4}$, agreeing with previously known results at order $S^{2}$.
We will see that this exponential form is extremely suitable for the
computation of the latter as a Fourier transform.

Our strategy is the following: we first consider the action of
the exponentiated soft factor acting on the three-point amplitude,
as an all-order extension of the Cachazo-Strominger soft theorem.
We have checked that this agrees with the known versions
of the Compton amplitude \cite{Arkani-Hamed:2017jhn,Bjerrum-Bohr:2017dxw} for $s\leq2$.
We leave the problem of obtaining the case $s\geq2$ for future investigation,
but we will comment on it at the end of section \ref{sec:factorization}.

To obtain \eqn{eq:compexp} we first propose
an all-order extension of the soft expansion~\eqref{eq:cachazostrominger} 
with respect to the graviton $k_4=\ket{4}[4|$:
\beal
\bigg[
\frac{(p_1\cdot\varepsilon_4)^{2}}{p_1 \cdot k_4}
\exp\!\bigg(i\frac{k_{4\mu}\varepsilon_{4\nu}J_1^{\mu\nu}}
                  {p_1\cdot\varepsilon_4}\bigg) +
\frac{(p_2\cdot\varepsilon_4)^{2}}{p_2 \cdot k_4}
\exp\!\bigg(i\frac{k_{4\mu}\varepsilon_{4\nu}J_2^{\mu\nu}}
                  {p_2\cdot\varepsilon_4}\bigg) & \\ +\,
\frac{(k_3\cdot\varepsilon_4)^{2}}{k_3 \cdot k_4}
\exp\!\bigg(i\frac{k_{4\mu}\varepsilon_{4\nu}J_3^{\mu\nu}}
                  {k_3\cdot\varepsilon_4}\bigg) &
\bigg] {\cal M}_3^{(s)}(p_1,p_2,k_3^+) .
\label{eq:expsoft}
\eeal
As stated in the introduction, two main problems arise when trying
to interpret \eqn{eq:cachazostrominger} as an exponential acting
on the lower-point amplitude. The first is that gauge invariance of
the denominator $p_i\cdot\varepsilon_4$ is not guaranteed.
Here we simply fix $\varepsilon_4^-=\sqrt{2}{\ket{4}[3|}/{[43]}$,
so the last term in \eqn{eq:expsoft} vanishes,
as we will show in a moment.
The second problem is that one still has to sum over two exponentials,
which would spoil the factorization of \eqn{eq:compexp}.
The solution is that
in this case both exponentials give the exact same contribution.
In the language of the previous section,
this is the fact that one can act with the operator
${k_{4\mu}\varepsilon_{4\nu}J^{\mu\nu}}/(p\cdot\varepsilon_4)$
either on $\bra{2}^{2s}$ or $\ket{1}^{2s}$,
giving the same result. 

Let us first inspect the three-point amplitude entering \eqn{eq:expsoft},
\be
{\cal M}_3^{(s)}(p_1,p_2,k_3^+) = {\cal M}_3^{(0)}
    \frac{\braket{12}^{2s}}{m^{2s}} , \qquad \quad
{\cal M}_3^{(0)} = m^2 x_3^2 = \frac{\bra{4}1|3]^2}{\braket{34}^2}
 = \frac{\braket{4|1|2|4}^2}{\braket{34}^4} ,
\label{eq:base3pt}
\ee
where we used $\varepsilon_3^+=\sqrt{2}{\ket{4}[3|}/{\braket{43}}$.
As explained in~\cite{Cachazo:2014fwa},
in order for the action of the differential operator to be well defined,
we need to solve momentum conservation and express ${\cal M}_3^{(0)}$
in terms of independent variables.
Solving for $|3]$ and $|4]$ yields the last expression in \eqn{eq:base3pt}.
Now to evaluate the third term in \eqn{eq:expsoft},
we recall from \app{app:angularmomentum}
\be
J_{3\;\!\alpha\dot{\alpha},\beta\dot{\beta}}^{\text{self-dual}}
  = 2i \la_{3(\alpha} \frac{\partial~}{\partial \la_3^{\beta)}}
    \epsilon_{\dot{\alpha}\dot{\beta}}
\qquad \Rightarrow \qquad
k_{4\mu} \varepsilon_{4\nu} J_3^{\mu\nu}
  =-\frac{i}{\sqrt{2}} \braket{43}
    \braket{4\frac{\partial}{\partial\la_3}}.
\ee
As the only place where $\bra{3}$ appears in \eqn{eq:base3pt}
is in the contraction with $\ket{4}$,
we see that the above differential operator annihilates
the scalar three-point amplitude ${\cal M}_3^{(0)}$.
Moreover, since the prefactor $\braket{12}^{2s}$
in the spin-$s$ amplitude ${\cal M}_3^{(s)}$
does not depend on $\ket{3}$,
we conclude that the exponential operator in the third term of \eqref{eq:expsoft} acts always trivially. The zeroth-order of the soft theorem $\propto (k_3\cdot\varepsilon_4)^{2}$ then vanishes by going to the chosen gauge, hence the last term drops as promised.

Let us now look at the angular momenta of the massive particles.
A similar inspection of
$\braket{4|1|2|4} = \braket{4\;\!1^a} [1_a 2_b] \braket{2^b 4}$
shows that the scalar piece ${\cal M}_{0}^{(3)}$
is in the kernel of the operators
\be
k_{4\mu} \varepsilon_{4\nu} J_1^{\mu\nu}
=-\frac{i}{\sqrt{2}} \braket{4\;\!1^a}
  \braket{4\;\!\frac{\partial~}{\partial \la_1^a}} , \qquad \quad
k_{4\mu} \varepsilon_{4\nu} J_2^{\mu\nu}
=-\frac{i}{\sqrt{2}} \braket{4\;\!2^a}
  \braket{4\;\!\frac{\partial~}{\partial \la_2^a}} .
\ee
Therefore, \eqn{eq:expsoft} is simplified to
\be
{\cal M}_3^{(0)}
\bigg[
\frac{(p_1\cdot\varepsilon_4)^{2}}{p_1 \cdot k_4}
\exp\!\bigg(i\frac{k_{4\mu}\varepsilon_{4\nu}J_1^{\mu\nu}}
                  {p_1\cdot\varepsilon_4}\bigg) +
\frac{(p_2\cdot\varepsilon_4)^{2}}{p_2 \cdot k_4}
\exp\!\bigg(i\frac{k_{4\mu}\varepsilon_{4\nu}J_2^{\mu\nu}}
                  {p_2\cdot\varepsilon_4}\bigg)
\bigg] \frac{\braket{12}^{2s}}{m^{2s}} .
\label{eq:expsoft2}
\ee
Moreover, our choice of the reference spinor for $\varepsilon_4$ implies
$p_1\cdot\varepsilon_4 = -p_2\cdot\varepsilon_4 = p\cdot\varepsilon$,
where $p = (p_1-p_2)/2$ is the average momentum
of the massive particle before and after Compton scattering.

From the discussion of the previous section
on the action of the angular-momentum operator
on $\bra{2}^{2s}$ and $\ket{1}^{2s}$,
we also have
\be
\exp\!\bigg(i\frac{k_{4\mu}\varepsilon_{4\nu}J_1^{\mu\nu}}
                  {p_1\cdot\varepsilon_4}\bigg) \braket{12}^{2s} =
\exp\!\bigg(i\frac{k_{4\mu}\varepsilon_{4\nu}J_2^{\mu\nu}}
                  {p_2\cdot\varepsilon_4}\bigg) \braket{12}^{2s} =
\bra{2}^{2s}
\exp\!\bigg(i\frac{k_{4\mu}\varepsilon_{4\nu}J^{\mu\nu}}
                  {p\cdot\varepsilon_4}\bigg) \ket{1}^{2s} .
\ee
Hence we obtain
\be
\frac{1}{m^{2s}}{\cal M}_3^{(0)}
\bigg[ \frac{(p_1\cdot\varepsilon_4)^2}{p_1 \cdot k_4}
     + \frac{(p_2\cdot\varepsilon_4)^2}{p_2 \cdot k_4} \bigg]
\bra{2}^{2s}
\exp\!\bigg(i\frac{k_{4\mu}\varepsilon_{4\nu}J^{\mu\nu}}
                  {p\cdot\varepsilon_4}\bigg) \ket{1}^{2s} ,
\ee
where we recognize the scalar Weinberg soft factor.
Recall that in this gauge $k_3\cdot\varepsilon_4=0$,
so there is no contribution from the other graviton.
As an easy check, we observe
that the scalar Compton amplitude, written \eg in~\cite{Arkani-Hamed:2017jhn,Bjerrum-Bohr:2017dxw},
can be constructed solely from this soft factor:
\be
{\cal M}_4^{(0)} =
{\cal M}_3^{(0)}
\bigg[ \frac{(p_1\cdot\varepsilon_4)^2}{p_1 \cdot k_4}
     + \frac{(p_2\cdot\varepsilon_4)^2}{p_2 \cdot k_4} \bigg]
 = -\frac{\bra{4}1|3]^4}{(2p_1\cdot k_4)(2p_2\cdot k_4)(2k_3 \cdot k_4)} .
\ee
This proves that \eqn{eq:compexp} can be obtained
from the all-order extension of the soft theorem~\eqref{eq:expsoft}.
Finally, the property~\eqref{eq:conj} is checked by
repeating the computation for the opposite-helicity graviton $k_3$.

\subsection{Factorization and soft theorems}
\label{sec:factorization}

In view of the exponentiation formulas,
we now show how factorization is realized in this operator framework. For the pole $(k_3+k_4)^2 \to 0$ it is evident,
so we will focus on the pole $(p_1 \cdot k_4) \to 0$.
In that limit the scalar part factors as
$ {\cal M}_4^{(0)} \to {{\cal M}_{3,\rm L}^{(0)} {\cal M}_{3,\rm R}^{(0)}} /
  {(2p_1 \cdot k_4)} $
corresponding to the product of the respective three-point amplitudes.
Let us denote the internal momentum by $p_{I}=p_1+k_4$.
Unitarity demands that the operator piece in \eqref{eq:compexp} behaves as
\be
\bra{2}^{2s}\!
\exp\!\bigg(i\frac{k_{4\mu}\varepsilon_{4\nu}J^{\mu\nu}}
                  {p\cdot\varepsilon_4}\bigg) \ket{1}^{2s}\!\to
\frac{1}{m^{2s}} [2|^{2s}\!
\exp\!\bigg(i\frac{k_{3\mu}\varepsilon_{3\nu}J^{\mu\nu}}
                  {p\cdot\varepsilon_3}\bigg) |I_a]^{2s} \bra{I^a}^{2s}\!
\exp\!\bigg(i\frac{k_{4\mu}\varepsilon_{4\nu}J^{\mu\nu}}
                  {p\cdot\varepsilon_4}\bigg) \ket{1}^{2s} .
\ee
Here the insertion of $p_{I}=|I_a] \bra{I^a}$ is needed
since the exponential operators act on different bases.
In order to show the above property,
it is enough to write the left factor in the chiral basis,
as in \sec{sec:exponentiation3pt},
which is possible on the three-particle kinematics
of the factorization channel:
\begin{align}
\frac{1}{m^{2s}} [2|^{2s}\!
\exp\!\bigg(i\frac{k_{3\mu}\varepsilon_{3\nu}J^{\mu\nu}}
                  {p\cdot\varepsilon_3}\bigg) |I_a]^{2s} & \bra{I^a}^{2s}\!
\exp\!\bigg(i\frac{k_{4\mu}\varepsilon_{4\nu}J^{\mu\nu}}
                  {p\cdot\varepsilon_4}\bigg) \ket{1}^{2s} \\ =
\frac{1}{m^{2s}} \braket{2\;\!I_a}^{2s} & \bra{I^a}^{2s}\!
\exp\!\bigg(i\frac{k_{4\mu}\varepsilon_{4\nu}J^{\mu\nu}}
                  {p\cdot\varepsilon_4}\bigg) \ket{1}^{2s} =
\bra{2}^{2s}\!
\exp\!\bigg(i\frac{k_{4\mu}\varepsilon_{4\nu}J^{\mu\nu}}
                  {p\cdot\varepsilon_4}\bigg) \ket{1}^{2s} . \nn
\end{align}
On the other hand,
we could have inserted the resolution of the identity
in the right factor
\beal
\frac{1}{m^{2s}} [2|^{2s}\!&
\exp\!\bigg(i\frac{k_{3\mu}\varepsilon_{3\nu}J^{\mu\nu}}
                  {p\cdot\varepsilon_3}\bigg) |I_a]^{2s} \bra{I^a}^{2s}\!
\exp\!\bigg(i\frac{k_{4\mu}\varepsilon_{4\nu}J^{\mu\nu}}
                  {p\cdot\varepsilon_4}\bigg) \ket{1}^{2s} \\ & =
\frac{1}{m^{2s}} [2|^{2s}\!
\exp\!\bigg(i\frac{k_{3\mu}\varepsilon_{3\nu}J^{\mu\nu}}
                  {p\cdot\varepsilon_3}\bigg) |I_a]^{2s} [I^a 1]^{2s} =
[2|^{2s}\!
\exp\!\bigg(i\frac{k_{3\mu}\varepsilon_{3\nu}J^{\mu\nu}}
                  {p\cdot\varepsilon_3}\bigg) |1]^{2s} .
\eeal
Putting this together with the scalar piece we can write, for instance,
\begin{align}
{\cal M}_4^{(s)} \xrightarrow[p_1 \cdot k_4 \to 0]{} &\,
\frac{{\cal M}_{3,\rm L}^{(0)} {\cal M}_{3,\rm R}^{(0)}}{2p_1 \cdot k_4}
\frac{1}{m^{2s}}
\bra{2}^{2s}
\exp\!\bigg(i\frac{k_{4\mu}\varepsilon_{4\nu}J^{\mu\nu}}
                  {p\cdot\varepsilon_4}\bigg) \ket{1}^{2s} \\ = &\,
\frac{{\cal M}_{3,\rm R}^{(0)}}{2p_1 \cdot k_4}
\exp\!\bigg(i\frac{k_{4\mu}\varepsilon_{4\nu}J_1^{\mu\nu}}
                  {p_1\cdot\varepsilon_4}\bigg)
{\cal M}_{3,\rm L}^{(0)} \frac{\braket{12}^{2s}}{m^{2s}} =
\frac{(p_1\cdot\varepsilon_4)^2}{p_1\cdot k_4}
\exp\!\bigg(i\frac{k_{4\mu}\varepsilon_{4\nu}J_1^{\mu\nu}}
                  {p_1\cdot\varepsilon_4}\bigg)
{\cal M}_{3,\rm L}^{(s)} . \nonumber
\end{align}
Here, using
${\cal M}_{3,\rm R}^{(0)}={\cal M}_3^{(0)}(p_1,p_I,k_4^-)
=2(p_1\cdot\varepsilon_4^-)^2$,
we have recovered the extension of the soft theorem~\eqref{eq:expsoft},
that we used as a starting point of this section,
in the limit $p_1 \cdot k_4 \rightarrow 0$.
The origin of the exponential soft factor in this case
is nothing but the three-point amplitude of spin-$s$ particles,
written as a series in the angular momentum.
Therefore, in our case the statement
of the subsubleading soft theorem~\eqref{eq:cachazostrominger}
follows from factorization of amplitudes of massive particles with spin.

Let us remark that, in analogy to the three-point case,
the exponential factor can be brought into a compact form
using identities like~\eqref{eq:raisinggeneral}.
For example, one can check that
\be
\bra{2}^{2s}
\exp\!\bigg(i\frac{k_{4\mu}\varepsilon_{4\nu}J_1^{\mu\nu}}
                  {p_1\cdot\varepsilon_4}\bigg)
\ket{1}^{2s} =
\bigg[ \braket{21} + \frac{[43]}{\bra{4}1|3]} \braket{24} \braket{41}
\bigg]^{2s}\! = m^{2s}
\bigg( \frac{[13]\braket{42} + \braket{14}[32]}{\bra{4}1|3]}
\bigg)^{\!2s} ,
\label{eq:comptonAHHderivation}
\ee
which converts the Compton amplitude into the form
\be
{\cal M}_4^{(s)} = 
-\frac{\bra{4}1|3]^{4-2s}}
      {(2p_1\cdot k_4)(2p_2\cdot k_4)(2k_3 \cdot k_4)} 
\big([13]\braket{42} + \braket{14}[32]\big)^{2s}
\label{eq:comptonAHH}
\ee
that is given in~\cite{Arkani-Hamed:2017jhn}.
We remark, however,
that this expression completely hides the spin dependence that
we need here for the classical computation.

It was pointed out in~\cite{Arkani-Hamed:2017jhn}
that the formula~\eqref{eq:comptonAHH} is only valid up to $s\leq2$.
For higher spins, one has to eliminate the spurious pole $\bra{4}1|3]$
that appears at the fifth order by adding contact terms.
From our perspective, this spurious pole corresponds precisely
to the contribution from $p_1\cdot\varepsilon_4$
appearing at higher orders in the soft expansion~\eqref{eq:comptonAHHderivation}.
Let us remark, however, that our result~\eqref{eq:compexp}
non-trivially extends the Cachazo-Strominger soft theorem
in the case of the Compton amplitude for minimally coupled spinning particles.
This is because for $s=2$ the exponential is truncated
only at the fourth order in the angular momentum,
whereas only the second order was guaranteed by the soft theorem.
This extension is what enables us in \sec{sec:HCL}
to obtain the scattering angle at order $S^{4}$,
by means of a Fourier transform acting directly on the exponential.
We leave the study of the contributions from contact terms
at higher spin orders for future work.

\section{Scattering angle as Leading Singularity}
\label{sec:HCL}

\subsection{Linearized stress-energy tensor of Kerr Solution}
\label{sec:energytensor}

In \sec{sec:multipole} we have shown that the
three-point and Compton amplitudes can be written in an exponential form.
We have also motivated the definition of a generalized expectation value
of an operator $\mathcal{O}$ acting on two massive states,
represented by their polarization tensors,
\be
   \braket{{\cal O}}
    = \frac{ \varepsilon_{2,\mu_1\ldots\mu_s}
             {\cal O}^{\mu_1\ldots\mu_s,\nu_1\ldots\nu_s}
             \varepsilon_{1,\nu_1\ldots\nu_s} }
           { \varepsilon_{2,\mu_1\ldots\mu_s} \varepsilon_1^{\mu_1\ldots\mu_s} } .
\ee

Let us first show how to apply this definition to match
the form of the stress-energy tensor of a single Kerr black hole
that we derived in the introduction:
\be
   h_{\mu\nu}(k) T^{\mu\nu}(-k)
    = (2\pi)^2 \delta(k^2)\delta(p \cdot k) (p\cdot\varepsilon)^2
      \exp\!\bigg(\!{-i}\frac{k_{\mu}\varepsilon_{\nu}S^{\mu\nu}}
                             {p\cdot\varepsilon}\bigg) ,
\label{eq:ht3}
\ee
There is a subtle but important point already present
in this classical matching that will guide us
in the following subsection on a path to the classical scattering angle.
The crucial difference between
the angular momentum operator $J^{\mu\nu}$ appearing in the soft theorem
and the classical spin $S^{\mu\nu}$ appearing in the expansion of $T^{\mu\nu}$
is that the latter satisfies the SSC~\eqref{eq:ssc}.
Moreover, there is an obvious sign flip in the respective exponents,
due to the sign difference between the differential and algebraic generators,
as mentioned in \sec{sec:spin1matter} and \app{app:angularmomentum}.
Therefore, following \sec{sec:spin1matter} (see also \app{app:spin1tensor})
we relate the two by
\be
   J^{\mu\nu} = -S^{\mu\nu}
      + \frac{1}{m^2} p^\mu p_\alpha J^{\alpha\nu}
      - \frac{1}{m^2} p^\nu p_\alpha J^{\alpha\mu} ,
\ee
which implies that the soft operator reads, at $p \cdot k=0$,
\be
   \frac{k_\mu \varepsilon_\nu J^{\mu\nu}}{p\cdot\varepsilon}
    = - \frac{k_\mu \varepsilon_\nu S^{\mu\nu}}{p\cdot\varepsilon}
      + \frac{1}{m^2} k_\mu p_\nu J^{\mu\nu} .
\label{eq:J2Sstart}
\ee

The key observation is that this operator acts on a chiral representation.
That is, for negative helicity,
if the states are built from the spinors $\ket{1}^{2s}$ and $\ket{2}^{2s}$
then the operator is algebraically realized by
$J^{\mu\nu}=-\sigma^{\mu\nu}=-i\sigma^{[\mu}\bar{\sigma}^{\nu]}/2$,
which is self-dual.
This means that 
\be
\frac{1}{m^2} k_\mu p_\nu J^{\mu\nu}
  = \frac{i}{2m^2} \epsilon^{\mu\nu\rho\sigma}
    k_\mu p_\nu J_{\rho\sigma}
  =-\frac{i}{2m^2} \epsilon^{\mu\nu\rho\sigma}
    k_\mu p_\nu S_{\rho\sigma}
  =-i a \cdot k .
\ee
On the three-point kinematics, one can show that
\be
   a \cdot k
    =  \pm i\frac{k_\mu \varepsilon_\nu^\pm S^{\mu\nu}}{p\cdot\varepsilon^\pm} ,
\label{eq:a2S}
\ee
so \eqn{eq:J2Sstart} becomes
\be
\label{eq:J2S}
   \frac{k_\mu \varepsilon_\nu J^{\mu\nu}}{p\cdot\varepsilon}
    = -2\frac{k_\mu \varepsilon_\nu S^{\mu\nu}}{p\cdot\varepsilon} .
\ee
It can be checked that
this factor-of-two relation is independent of the helicity of the graviton.
To compute the generalized expectation value, we will also need to
consider the product $\varepsilon_{1}^{(s)}\cdot\varepsilon_{2}^{(s)}$.
To that end we use the following representation of polarization
tensors, obtained as tensor products of the spin-1 polarization vectors~\eqref{eq:polvectorsmassive}
\be
\varepsilon_{1}^{(s)}=\varepsilon_{1}^{\otimes s}
=\frac{2^{s/2}}{m^{s}}\big(|1\rangle[1|\big)^{\odot s} , \qquad \quad 
\varepsilon_{2}^{(s)}=\varepsilon_{2}^{\otimes s}
=\frac{2^{s/2}}{m^{s}}\big(|2\rangle[2|\big)^{\odot s} ,
\ee
where we now take $p_2$ to be outgoing,
so $\ket{2}$ is minus that of \sec{sec:multipole}.
This leads to
\beal
\lim_{s\to\infty} m^{2s}
\varepsilon_{2,\mu_1\ldots\mu_s} \varepsilon_1^{\mu_1\ldots\mu_s}
= \lim_{s\to\infty} \braket{21}^s [12]^s &
= \lim_{s\to\infty} \bra{2}^{2s}
  \bigg(1+\frac{\ket{k} \bra{k}}{m x}\bigg)^{\!s} \ket{1}^{2s} \\
= \lim_{s\to\infty} \bra{2}^{2s}
  \Bigg[ \sum_{j=0}^{s} \binom{s}{j}
       \bigg( \frac{\ket{k}\bra{k}}{m x}\bigg)^j \Bigg] \ket{1}^{2s} &
= \lim_{s\to\infty} \bra{2}^{2s}
  \Bigg[ \sum_{j=0}^{s} \binom{2s}{j}
       \bigg( \frac{\ket{k}\bra{k}}{2m x}\bigg)^j \Bigg] \ket{1}^{2s} \\
= \lim_{s\to\infty} \bra{2}^{2s}
  \exp\!\bigg(\frac{i}{2}\frac{k_\mu \varepsilon_\nu J^{\mu\nu}}
                              {p\cdot\varepsilon}\bigg) \ket{1}^{2s} &
= \lim_{s\to\infty}
  \exp\!\bigg(\!{-i}\frac{k_\mu \varepsilon_\nu S^{\mu\nu}}
                         {p\cdot\varepsilon}\bigg) \braket{21}^{2s} ,
\label{eq:norm}
\eeal
where we have used the $s\to \infty$ limit of \eqref{eq:raisinggeneral2} and in the last line we extracted the operator as a GEV.
The same manipulation can be done
for the three-point minus-helicity amplitude:
\be
\lim_{s\to\infty} m^{2s} \varepsilon_{2,\mu_1\ldots\mu_s}
   {\cal M}_3^{(s),\mu_1\ldots\mu_s,\nu_1\ldots\nu_s}
   \varepsilon_{1,\nu_1\ldots\nu_s}
    = m^2 x^2 \lim_{s\to\infty}
      \exp\!\bigg(\!{-2i}\frac{k_{\mu}\varepsilon_{\nu}S^{\mu\nu}}{p\cdot\varepsilon}
            \bigg) \braket{21}^{2s} .
\ee
Here we would like to emphasize a key point.
Even though the exponential operator is always present at finite spin,
it is only in the infinite-spin limit that the expansion does not truncate.
This leads to
\be
\lim_{s\to\infty} \braket{{\cal M}_3^{(s)}} = 2(p\cdot\varepsilon)^2
\exp\!\bigg(\!{-i}\frac{k_\mu \varepsilon_\nu S^{\mu\nu}}{p\cdot\varepsilon}\bigg) ,
\label{eq:GEV3}
\ee
which recovers the Kerr gravitational coupling~\eqref{eq:ht3},
as promised in \eqn{eq:vertex2amp3}, ---
this time with the SSC condition incorporated.
The plus-helicity graviton gives the same GEV.
One can also keep the minus helicity
and redo the computation in the antichiral basis:
\begin{subequations}
\begin{align} &\!\!\!\!
\lim_{s\to\infty} m^{2s} \varepsilon_{2,\mu_1\ldots\mu_s}
   {\cal M}_3^{(s),\mu_1\ldots\mu_s,\nu_1\ldots\nu_s}
   \varepsilon_{1,\nu_1\ldots\nu_s}
    = m^2 x^2 \lim_{s\to\infty} [21]^{2s} , \\ &\!\!\!\!
\lim_{s\to\infty} m^{2s}
\varepsilon_{2,\mu_1\ldots\mu_s} \varepsilon_1^{\mu_1\ldots\mu_s}
    = \lim_{s\to\infty}
      \exp\!\bigg(\!{-i}\frac{k_\mu \varepsilon_\nu^+ S^{\mu\nu}}
                             {p\cdot\varepsilon^+}
      \bigg) [21]^{2s}
    = \lim_{s\to\infty}
      \exp\!\bigg(i\frac{k_\mu \varepsilon_\nu^- S^{\mu\nu}}{p\cdot\varepsilon^-}
      \bigg) [21]^{2s} .\!\!
\end{align}
\end{subequations}
Therefore, the GEV~\eqref{eq:GEV3} is invariant
with respect to the choice of the spinor basis as well.

Finally, we notice that the self-dual condition is natural
when considering a definite-helicity coupling,
\eg $k_\mu \varepsilon^-_\nu J^{\mu\nu}$
projects out the anti-self-dual piece.
However, we should keep in mind that
this is just an artifact of our choice of chiral spinor basis
to describe that coupling.
It would be interesting to find a non-chiral form,
analogous to the vector parametrization of \sec{sec:spin1matter},
in such a way that the amplitude already contains
the covariant-SSC spin tensor built in.

\subsection{Kinematics and scattering angle for aligned spins}
\label{sec:alignedspin}

We now consider scattering of two massive spinning particles,
one with mass $m_a$, spin (quantum number) $s_a$,
initial momentum $p_1$, and final momentum $p_2$,
and the other with mass $m_b$, spin $s_b$,
initial momentum $p_3$, and final momentum $p_4$,
\be
p_1^2=p_2^2=m_a^2, \qquad \quad
p_3^2=p_4^2=m_b^2,
\ee
following here the conventions of \cite{Guevara:2017csg}.
The total amplitude
\be\label{M2BH}
\mc M_4^{(s_a,s_b)}\;\;=\;\;
\begin{array}{c}
p_2 \\ \\ \\ p_1
\end{array}
\vcenter{\hbox{\includegraphics[scale=.25]{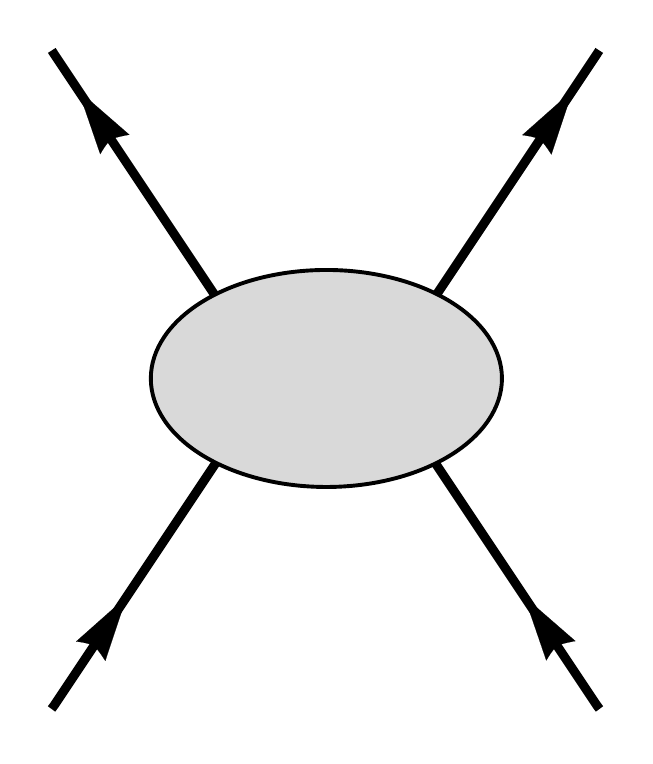}}}
\begin{array}{c}
p_4 \\ \\ \\ p_3
\end{array}
\ee
is a function of the external momenta
and the external spin states (polarization tensors). 
We define as usual
\be
s=p_{\rm tot}^2,  \qquad \quad
t=k^2,
\ee
where $p_\mr{tot}$ is the total momentum, and $k$ is the momentum transfer,
\be
p_{\rm tot}=p_1+p_3=p_2+p_4 , \qquad \quad
k=p_2-p_1=p_3-p_4 .
\ee
The Mandelstam variable $s$, the total center-of-mass-frame energy $E$,
the relative velocity $v$ (between the inertial frames
attached to the incoming momenta $p_1$ and $p_3$, with $v>0$),
and the corresponding relative Lorentz factor $\gamma$
--- each of which determines all the others,
given fixed rest masses $m_a$ and $m_b$ --- are related by
\be\label{eq:sE}
s = E^2 = m_a^2 + m_b^2 + 2m_a m_b \gamma , \qquad\quad
\frac{p_1 \cdot p_3}{m_a m_b} = \gamma = \frac{1}{\sqrt{1-v^2}} .
\ee

At $t=0$, it is convenient to fix the little-group scaling
of the internal graviton (for tree-level one-graviton exchange).
Following \cite{Guevara:2017csg}, we can choose it as 
\be\label{eq:x3}
x_b = \sqrt{2}\,\frac{p_b\cdot\varepsilon^-(-k)}{m_b}
    =-\sqrt{2}\,\frac{p_b\cdot\varepsilon^-\!}{m_b} = 1 .
\ee
This implies
\be\label{eq:x1}\!
x_a^{-1} = -\sqrt{2}\,\frac{p_a\cdot\varepsilon^+\!}{m_a}
   = -\frac{\bra{r}p_a|k]}{m_a \braket{r\;\!k}}
   = \gamma (1-v) , \qquad \quad
x_a = \sqrt{2}\,\frac{p_a\cdot\varepsilon^-\!}{m_a}
   = -\frac{[r|p_a\ket{k}}{m_a [r\;\!k]}
   = \gamma (1+v) .
\ee

We consider the case, in the classical limit, in which the two particles' rescaled spin vectors
\be
a_a^\mu=\frac{1}{2m_a^2}\varepsilon^\mu{}_{\nu\rho\sigma}p_a^\nu S_a^{\rho\sigma},
\qquad \quad
a_b^\mu=\frac{1}{2m_b^2}\varepsilon^\mu{}_{\nu\rho\sigma}p_b^\nu S_b^{\rho\sigma},
\label{eq:aaab}
\ee
are aligned with the system's total angular momentum. They are orthogonal to the constant scattering plane, and are conserved. The scattering plane is defined containing all the momenta, see e.g. \cite{Vines:2017hyw}. Here $p_a$ is the average momentum $p_a=(p_1+p_2)/2=p_1+\mc O(k)=p_2+O(k)$, similarly for $p_b$.  In this ``aligned-spin case", up to order $G^2$,  we will find that the classical scattering angle $\theta$ by which both bodies are scattered in the center-of-mass frame, is given by the same relation as for the spinless case \cite{Kabat:1992tb,Akhoury:2013yua,Bjerrum-Bohr:2018xdl},
\be
\theta + {\cal O}(\theta^3)
 = 2\sin\frac{\theta}{2}
 = -\frac{E}{(2m_a m_b \gamma v)^2}
   \frac{\partial~}{\partial b} \int\!\frac{d^2\bs k}{(2\pi)^2}\,
   e^{i\bs k\cdot\bs b}\!\lim_{s_a,s_b \to \infty}
   \braket{{\cal M}_4^{(s_a,s_b)}} + {\cal O}(G^3) ,
\label{eq:angle}
\ee
where $\braket{{\cal M}_4^{(s_a,s_b)}}$
is the generalized expectation value of the amplitude~\eqref{M2BH},
the momentum transfer $\bs k$ is integrated over the 2D scattering plane,
and $\bs b$ is the vectorial impact parameter with magnitude~$b$,
counted from the second particle to the first as in \cite{Vines:2017hyw}.  Compared to the nonspinning/scalar case, this version of \eqref{eq:angle} differs only in that the aligned spin components $a_a$ and $a_b$, the magnitudes of the vectors in \eqref{eq:aaab}, will appear as scalar parameters in the amplitude.  While we do not claim to provide a first-principles derivation of the applicability of \eqref{eq:angle} to the spinning case with aligned spins, we find that its use here produces results which are (quite nontrivially) fully consistent with the results of \cite{Bini:2017xzy,Vines:2017hyw,Bini:2018ywr,Vines:2018gqi} for aligned-spin scattering angles for binary black holes.

\subsection{First post-Minkowskian order}
\label{sec:1PM}

\begin{figure}[t]
\centering
\vspace{-10pt}
\includegraphics[width=0.4\textwidth]{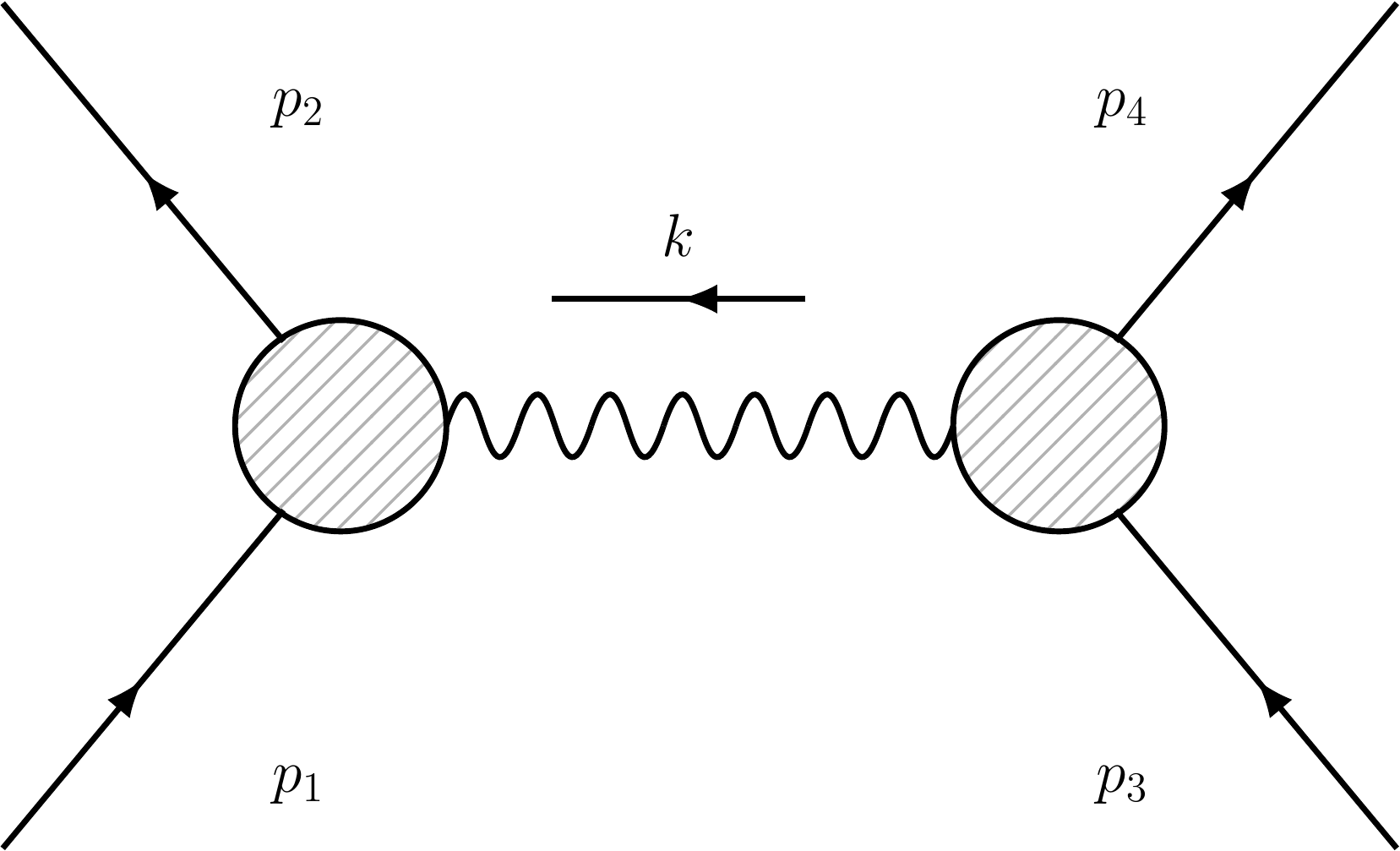}
\vspace{-5pt}
\caption{\small Tree-level singularity for one-graviton exchange}
\label{fig:1PM}
\end{figure}

At 1PM or tree level, the leading-singularity prescription
reduces to a $t$-channel residue
equivalent to one-graviton exchange~\cite{Cachazo:2017jef}.
The reason that this leads to classical effects is that the $O(t^0)$ piece,
which is dropped,
is ultralocal after a Fourier transform \cite{Donoghue:1994dn,Neill:2013wsa}.
In contrast to the one-loop case,
the HCL defined as the leading order in $t$
is trivially implemented from the fact that
the computation is done under the support of the factorization channel.
Following sections 3.1 and 4.2 of \cite{Guevara:2017csg},
the LS for the amplitude~\eqref{M2BH} with one graviton exchange
is obtained by gluing two massive higher-spin three-point amplitudes
at minimal coupling, see \fig{fig:1PM}.
These amplitudes are now given in the exponential form
by \eqns{eq:exp3m}{eq:exp3p} in the chiral basis.
Summing over helicities, we have
\beal
\hat{\cal M}_4^{(s_a,s_b)} & = \frac{1}{t} \bigg[
\hat{\cal M}_3^{(s_a)}(p_1,-p_2,k^-) \otimes
\hat{\cal M}_3^{(s_b)}(p_3,-p_4,-k^+) \\ & \qquad~ +
\hat{\cal M}_3^{(s_a)}(p_1,-p_2,k^+) \otimes
\hat{\cal M}_3^{(s_b)}(p_3,-p_4,-k^-) \bigg] \\ & = 
\frac{m_a^2m_b^2}{t}
\bigg[ \frac{x_a^2}{x_b^2}
       \exp\!\bigg(i\frac{k_\mu\varepsilon^-_\nu J^{\mu\nu}_a}
                         {p_a\cdot\varepsilon^-}\bigg)
     + \frac{x_b^2}{x_a^2}
       \exp\!\bigg({-i}\frac{k_\mu\varepsilon^-_\nu J^{\mu\nu}_b}
                            {p_b\cdot\varepsilon^-}\bigg) \bigg] .
\eeal
Here we will take the limit where both massive particles' spin quantum numbers
($s_a$ and~$s_b$) go to infinity.
After using \eqns{eq:a2S}{eq:J2S} valid on the three-point kinematics,
we can rewrite the exponents in a form
independent of the polarization vector:
\begin{subequations}
\begin{align}
   +i\frac{k_\mu \varepsilon^-_\nu J^{\mu\nu}_a}{p_a\cdot\varepsilon^-} &
    = -2i \frac{k_\mu\varepsilon^-_\nu S^{\mu\nu}_a}{p_a\cdot\varepsilon^-}
    = +2a_a \cdot k
    = +2i \epsilon_{\mu\nu\rho\sigma}
      \frac{p_a^\mu p_b^\nu k^\rho a_a^\sigma}{m_a m_b\gamma v}
    = +2i\bs k\times\hat{\bs p}\cdot \bs a_a , \\
   {-i}\frac{k_\mu \varepsilon^-_\nu J^{\mu\nu}_b}{p_b\cdot\varepsilon^-} &
    = +2i \frac{k_\mu \varepsilon^-_\nu S^{\mu\nu}_b}{p_b\cdot\varepsilon^-}
    = -2a_b \cdot k
    = -2i \epsilon_{\mu\nu\rho\sigma}
      \frac{p_a^\mu p_b^\nu k^\rho a_b^\sigma}{m_a m_b\gamma v}
    = -2i\bs k\times\hat{\bs p}\cdot \bs a_b .
\end{align}
\end{subequations}
Here we used the on-shell equality
$ i \epsilon_{\mu\nu\rho\sigma} p_a^\mu p_b^\nu k^\rho a^\sigma
  = m_a m_b \sqrt{\gamma^2-1}\:\!(k \cdot a) $
to reintroduce the Levi-Civita tensor
and thus to expose the scalar triple products in the center-of-mass frame,
where $\hat{\bs p}$ is the unit vector in the direction of the relative momentum.
Moreover, recall that on the three-point helicity factors satisfy
the seemingly contradictory conditions
$x_a/x_b = \gamma(1+v)$ and $x_b/x_a = \gamma(1-v)$,
as indicated by \eqns{eq:x3}{eq:x1}.
Finally, restoring the prefactor of $-(\kappa/2)^2=-8\pi G$,
and dividing by the normalization factor arising from the generalized expectation value as in \eqn{eq:norm}, 
\be
   (\varepsilon_1\cdot\varepsilon_2)(\varepsilon_3\cdot\varepsilon_4)
   ~\rightarrow~\exp\!\Big(i\bs k\times\hat{\bs p}\cdot(\bs a_a-\bs a_b)\Big)
\label{GEVnorm}
\ee
(with the relative sign due to the direction of $k$),
we obtain
\be
\braket{{\cal M}_4} = 8\pi G\frac{m_a^2m_b^2}{-t} \gamma^2
   \sum_{\pm} (1\pm v)^2
   \exp\!\Big({\pm i}\bs k\times\hat{\bs p}\cdot(\bs a_a + \bs a_b)\Big) .
\ee
Inserting this into the scattering-angle formula~\eqref{eq:angle} gives
\beal
   \theta_\text{tree} & = -\frac{GE}{v^2} \sum_\pm (1 \pm v)^2
      \frac{\partial~}{\partial b} \int\!\frac{d^2\bs k}{2\pi\bs k^2}
      \exp\!\Big( i\bs k \cdot \big[ \bs b \pm \hat{\bs p}\times(\bs a_a + \bs a_b)
                               \big] \Big) \\ &
    = \frac{GE}{v^2} \sum_\pm (1 \pm v)^2
      \frac{\partial~}{\partial b}
      \bigg[ \log\big|\bs b\pm\hat{\bs p}\times(\bs a_a+\bs a_b)\big|
           = \log\big(b\pm(a_a+a_b)\big) \bigg] \\ &
    = \frac{GE}{v^2}\sum_\pm\frac{(1\pm v)^2}{b\pm(a_a+a_b)} ,
\label{thetatreeh}
\eeal
having used $\hat{\bs p}\times \bs a=a\bs b/b$ for both spins in the aligned-spin configuration.  This precisely matches the result for the 1PM aligned-spin binary-black-hole scattering angle found in~\cite{Vines:2017hyw}.

Finally, let us emphasize that, as stated in the introduction,
this already differs from the strategy implemented
in \eg \cite{Holstein:2008sx,Vaidya:2014kza},
where the full tree-level amplitude for $s=\{\frac{1}{2},1,2\}$
was computed in the first place.
Only then it was expanded in the NR limit $k=(0,\bs k)\to 0$
under the COM frame.
The evaluation of spin effects requires tracking subleading orders
in the momentum transfer $\bs k$ (denoted there by $\!\vec{\,q}$),
which in general contain both classical and quantum pieces,
depending on whether they include the corresponding power of the spin vector.
This is precisely what the LS singles out
by dropping the (quantum) contraction $t=k^2$
in favor of the (classical) tensor structures $\sim k^n S^n$.
At tree level this is equivalent to set the HCL $t=0$,
but at one loop the HCL is needed to drop further quantum contributions
from the LS, as we shall explain in the next subsection.

\subsection{Second post-Minkowskian order}
\label{sec:2PM}

\begin{figure}[t]
\centering
\vspace{-10pt}
\includegraphics[width=0.4\textwidth]{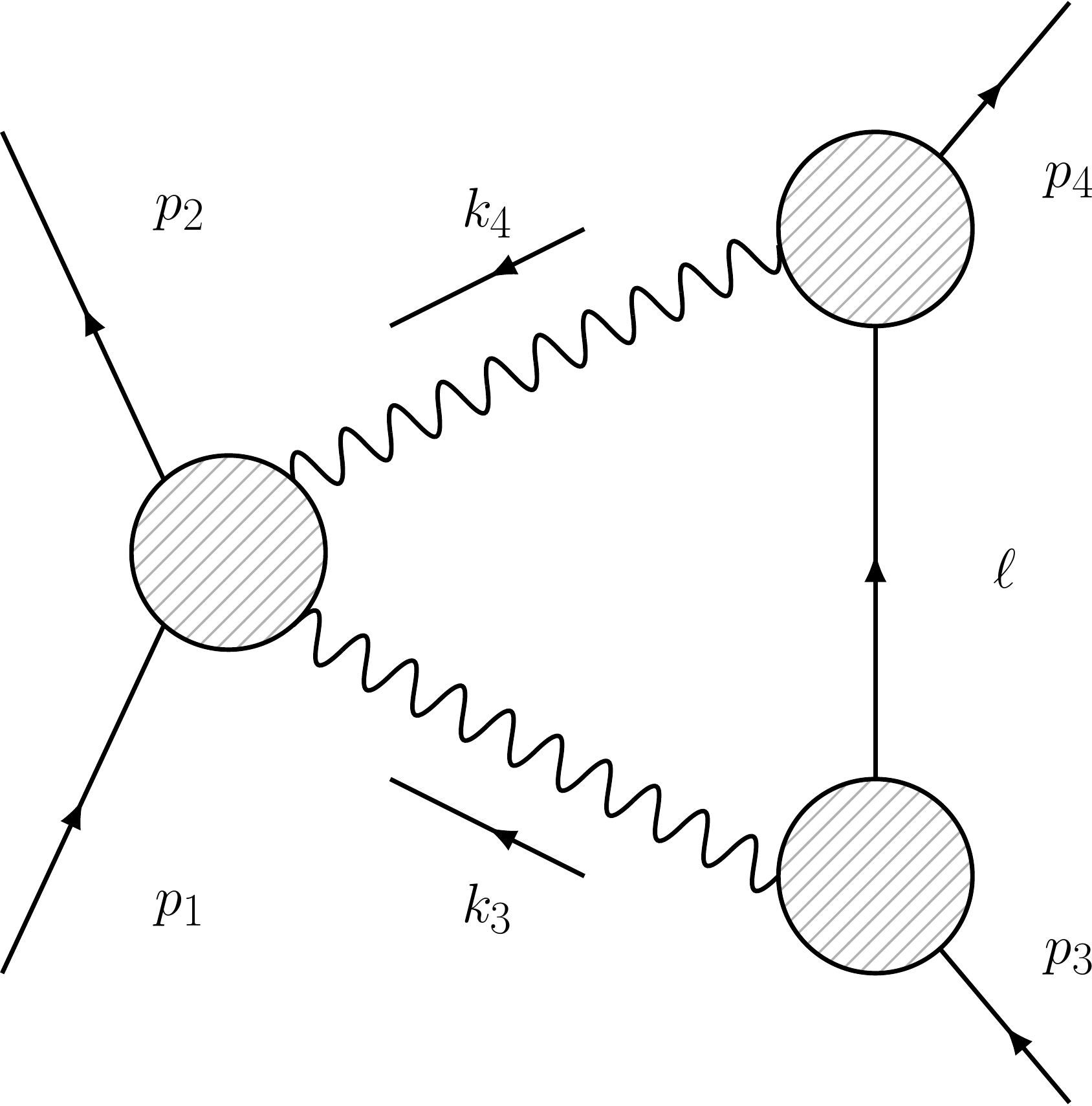}
\vspace{-5pt}
\caption{\small Triangle leading-singularity configuration}
\label{fig:triangle}
\end{figure}

In this section we derive a compact form for the 2PM (or $\mathcal{O}(G^2)$) aligned-spin scattering angle. It is obtained from the one-loop version of the four-point amplitude~\eqref{M2BH} through the triangle leading singularity proposed in~\cite{Guevara:2017csg} for computing its classical piece. The LS is now given by a contour integral for a single complex variable $y$ that remains in the loop integration after cutting the three propagators of \fig{fig:triangle}:
\be
    \ell^2(y)=m_b^2, \qquad (p_3-\ell(y))^2=0, \qquad (p_4-\ell(y))^2=0 .
\ee
It was argued in~\cite{Bjerrum-Bohr:2013bxa,Cachazo:2017jef,Guevara:2017csg} that for the spinless case the Compton amplitude for identical helicities leads to no classical contribution. This fact is also true for arbitrary spin, as will be proven somewhere else. This implies that only the opposite-helicity case treated in \sec{sec:exponentiation4pt} is needed, together with three-point interactions. The derivation is thus valid (to describe minimally coupled elementary particles) at least up to $\mathcal{O}(a_a^4)$ and to all orders in $a_b$, where $a_a$ is the rescaled spin of the particle that appears in the Compton amplitude, and $a_b$ is the spin of other particle. As explained already in~\cite{Arkani-Hamed:2017jhn,Guevara:2017csg} and emphasized in \sec{sec:exponentiation4pt} the Compton amplitude needs the introduction of contact terms for $s_a>2$. Nevertheless, the exponential structure found already for $s_a\leq2$ fits very nicely into the Fourier transform and leads to a compact formula for the scattering function, which can be computed directly once the multipole operators have been identified. The final formula resums all orders in both spins, but is not justified starting at $\mathcal O(a_a^5)$.  We finally expand in spins and find perfect agreement with the linear- and quadratic-order-in-spin results of  \cite{Bini:2018ywr} and \cite{Vines:2018gqi}. The computation of the possible contributions to the LS from contact terms arising in the higher-spin Compton amplitude is left for future work.

Our strategy is to identify the spin-multipole-coupling operators  $\bs k\times\hat{\bs p}\cdot \bs a_a$ and $\bs k\times\hat{\bs p}\cdot \bs a_b$ in the exponential form of the three and four point amplitudes entering the triangle leading singularity, see \fig{fig:triangle}.
This is done on the support of the Holomorphic Classical Limit,\footnote{The name ``Holomorphic Classical Limit'' is due to the external momenta being complex at that point.
}
which accounts for a null momentum transfer $k^2=0$ and recovers the three-point kinematics studied in \sec{sec:multipole}. The soft expansion in $k$ accounts for a simultaneous expansion in both powers of spin. 

Let us first recap the triangle leading singularity,
also introducing a more economic formulation of it.
It consists of a contour integral obtained by gluing three-point amplitudes with the Compton amplitude. Our starting point is the expression 
\be\label{eq:aS}
\frac{i(\kappa/2)^4}{8m_b\sqrt{-t}}
\int_{\Gamma_{\rm{LS}}}\!\frac{dy}{2\pi y}
\hat{\cal M}_4^{(s_a)}(p_1,-p_2,k_3^+,k_4^-) \otimes
\hat{\cal M}_3^{(s_b)}(p_3,-\ell,-k_3^-)
\frac{\ket{\ell}^{2s} \bra{\ell}^{2s}\!}{m_b^{2s}}
\hat{\cal M}_3^{(s_b)}(-p_4,\ell,-k_4^+) ,
\ee
where we have inserted the operator $\ket{\ell} \bra{\ell}$
in-between the three-point amplitudes to denote operator multiplication,
in the same sense as in \sec{sec:spin1matter}.
Here $\Gamma_{\rm{LS}}$ is the leading-singularity contour
that can be obtained at either $|y|=\epsilon$ or $|y|\rightarrow \infty$.
The loop momenta, together with their corresponding
spinors, are functions of $y$ given by eq.~(3.17) of \cite{Guevara:2017csg}.
Here we will only need the following limits:
\begin{align}
|k_3] & = \frac{1}{2} |k](1+y)
+ {\cal O}\bigg(\frac{\sqrt{-t}}{m_{b}}\bigg) , \qquad \quad
\bra{k_3} = \frac{1}{2y} \bra{k}(1+y)
+ {\cal O}\bigg(\frac{\sqrt{-t}}{m_{b}}\bigg) , \nn \\
|k_4] & = \frac{1}{2} |k](1-y)
+ {\cal O}\bigg(\frac{\sqrt{-t}}{m_{b}}\bigg) , \qquad \quad
\bra{k_4} =-\frac{1}{2y} \bra{k}(1-y)
+ {\cal O}\bigg(\frac{\sqrt{-t}}{m_{b}}\bigg) , \label{limits} \\
\braket{k_3 k_4} & = \frac{\sqrt{-t}}{y}
+ {\cal O}\bigg(\frac{t}{m_b^2}\bigg) , \qquad\!\qquad
\bra{k_4}p_1|k_3] = \frac{m_a \gamma}{2y} [2y-v(1+y^2)] \sqrt{-t}
+ {\cal O}\bigg(\frac{t}{m_b^2}\bigg) . \nn
\end{align}

Recall that at $t=0$ the momentum transfer reads $k=|k]\bra{k}$
and the scaling of the spinors $|k]$, $\bra{k}$ is
fixed by the condition~\eqref{eq:x3}.
In turn, this fixes the little-group scaling of
both internal gravitons $k_3$ and $k_4$.
We can now insert the exponential expressions (for $s_a \leq 2$)
and evaluate the scalar pieces, obtaining
\beal &
\frac{i(\kappa/2)^4}{8m_b\sqrt{-t}}\!\int_{\Gamma_{\rm{LS}}}\!
    \frac{dy}{2\pi y} {\cal M}_4^{(0)}(p_1,-p_{2},k_{3}^{+},k_{4}^-)
    {\cal M}_3^{(0)}(p_{3},-{\ell},-k_{3}^{-})
    {\cal M}_3^{(0)}(-p_{4},{\ell},-k_{4}^{+}) \\ & \qquad \qquad\!\quad \times
    \exp\!\bigg(i\frac{k_{4\mu}\varepsilon_{4\nu}^-J_a^{\mu\nu}}
                      {p_1\cdot\varepsilon_4^-}\bigg)\otimes
    \exp\!\bigg({-i}\frac{k_{3\mu}\varepsilon_{3\nu}^-J_b^{\mu\nu}}
                         {p_3\cdot\varepsilon_3^-}\bigg)  \\ &
 =-\frac{i\kappa^4 m_a^2 m_b^3 \gamma^2}{2^9 v^2 \sqrt{-t}}\!
   \int_{\Gamma_{\rm{LS}}}\!\!
   \frac{dy\,[2y-v(1+y^2)]^{4}}{2\pi y^3(1-y^2)^2}
   \exp\!\bigg(i\frac{k_{4\mu}\varepsilon_{4\nu}^-J_a^{\mu\nu}}
                     {p_1\cdot\varepsilon_4^-}\bigg)\otimes
   \exp\!\bigg({-i}\frac{k_{3\mu}\varepsilon_{3\nu}^-J_b^{\mu\nu}}
                        {p_3\cdot\varepsilon_3^-}\bigg) ,
\eeal
to leading orders in $t$.

Before proceeding to compute the GEV, let us clarify an important point. Recall that in the tree-level case the exponential operator
was truncated at order $2s$ in the expansion. The infinite spin limit
did not alter the lower orders in the exponential but simply accounted
for promoting such finite number of terms to a full series. We assume such condition still holds for the Compton amplitude, that is, the first five orders reproducing the exponential expansion are not spoiled in the infinite spin limit. The reason is that at arbitrary spin, the introduction of contact terms is only needed to cancel the spurious pole coming from the exponent, which appears as a pole in the amplitude only at fifth order.

With the previous consideration, the above operator formula in the infinite spin limit is fourth-order exact in the expansion of the left exponential and fully exact in the expansion of the right exponential. Let us now proceed to evaluate the exponents of both. The exponential factor on
the right can be obtained straight at $t=0$ kinematics. In fact,
using 
\be
k_3 = \frac{(1+y)^2}{4y}k ,
\ee
we find
\be
\exp\!\bigg({-i}\frac{k_{3\mu}\varepsilon_{3\nu}^-J_b^{\mu\nu}}
                     {p_3\cdot\varepsilon_3^-}\bigg)
=\exp\!\bigg({-i}\frac{(1+y)^2}{4y}
                 \frac{k_\mu\varepsilon_{\nu}^-J_b^{\mu\nu}}
                      {p_3\cdot\varepsilon^-}\bigg)
=\exp\!\bigg({-i}\frac{(1+y)^2}{2y} \bs k\times\hat{\bs p}\cdot\bs a_b\bigg),
\ee
where the polarization vector $\varepsilon_3^-$ for $k_3$
can be taken as the vector~$\varepsilon^-$ for $k$,
up to a scale that cancels. We have again identified
$k_{\mu}\varepsilon_{\nu}J_{b}^{\mu\nu}/(p_3\cdot\varepsilon_3)
=2\bs k\times\hat{\bs p}\cdot\bs a_b$
as the classical operator that will enter the GEV, whereas the $y$
dependence contributes to the contour integral.

Now, recall that the
left exponential corresponds to the Compton amplitude and was fixed
in \sec{sec:exponentiation4pt} using $k_3\cdot\varepsilon_4=0$, \ie
\be
\varepsilon_4^-=-\sqrt{2}\frac{|k_3]\bra{k_4}}{[k_3 k_4]},
\ee
which is singular at $t=0$.
In order to evaluate, it we will need the following trick.
First note that at $t\neq0$ the numerator is gauge invariant,
hence we can write
\be
k_{4\mu}\varepsilon_{4\nu}^- J_a^{\mu\nu}
=k_{4\mu}\hat{\varepsilon}_{4\nu}^- J_a^{\mu\nu} ,
\ee
where 
\be
\hat{\varepsilon}_4^- = -\sqrt{2}\frac{|r]\bra{k_4}}{[r\;\!k_4]}
\ee
and $|r]$ is some reference spinor such that $[r\;\!k_4] \neq 0$.
This means that in the limit we have
\begin{eqnarray}\label{eq:lim}
\lim_{t \rightarrow 0}
\frac{k_{4\mu}\varepsilon_{4\nu}^- J_a^{\mu\nu}}{p_1\cdot\varepsilon_4^-}
&=& \big(k_{4\mu} \hat{\varepsilon}_{4\nu}^- J_a^{\mu\nu}\big)_{t=0}
  \lim_{t\rightarrow0} (p_1\cdot\varepsilon_4^-)^{-1} \nonumber \\
&=&\bigg(\frac{k_{4\mu} \hat{\varepsilon}_{4\nu}^- J_a^{\mu\nu}} {p_1\cdot\hat{\varepsilon}_4^-}\bigg)_{t=0}\times  (p_1\cdot\hat{\varepsilon}_4^-)\big|_{t=0}  
  \lim_{t\rightarrow0} (p_1\cdot\varepsilon_4^-)^{-1} \,. \end{eqnarray}
The limit can be evaluated directly using \eqn{limits}. We find
\be
\lim_{t \rightarrow 0} (p_1\cdot\varepsilon_4^-)
=-\frac{\gamma m_a}{2\sqrt{2} y^2} \big[2y-v(1+y^2)\big] .
\ee
On the other hand, recall that at $t=0$ we recover
three-particle kinematics for $p_1$, $p_2$ and $k$.
This means that the combination 
\be
(p_1\cdot\hat{\varepsilon}_4^-)\big|_{t=0}
=-\frac{[r|p_1\ket{k_4}}{\sqrt{2}[r\;\!k_4]} \bigg|_{t=0}
=+\frac{1}{y} \frac{[r|p_1\ket{k}}{\sqrt{2}[r\;\!k]}
\ee
is independent of the choice of $r$. Using \eqn{eq:x1} we can identify this factor with
\be
{-}\frac{1}{y} (p_1\cdot\varepsilon^-) = -\frac{\gamma m_a}{\sqrt{2}y}(1+v) ,
\ee
Putting all together in \eqref{eq:lim} and using $k_4=-\frac{(1-y)^2}{4y}k$,
we have 
\beal
\lim_{t \rightarrow 0}
\frac{k_{4\mu} \varepsilon_{4\nu}^- J_a^{\mu\nu}}{p_1\cdot\varepsilon_4^-}&
= \bigg(\frac{k_{4\mu} \hat{\varepsilon}_{4\nu}^- J_a^{\mu\nu}}
             {p_1\cdot\hat{\varepsilon}_4^-}\bigg)_{t=0}
  \times \frac{1}{y} (p_1\cdot\varepsilon^-) \times
  \frac{2\sqrt{2}y^2}{\gamma m_a} \big[2y-v(1+y^2)\big]^{-1} \\ &
=-\frac{(1-y)^2(1+v)}{4y-2v(1+y^2)}
  \bigg(\frac{k_{\mu} \varepsilon_{\nu}^- J_a^{\mu\nu}}
             {p_1\cdot\varepsilon^-}\bigg)
=-\frac{(1-y)^2(1+v)}{2y-v(1+y^2)} \bs k\times\hat{\bs p}\cdot\bs a_a .
\eeal

Attaching the same normalization~\eqref{GEVnorm}
as in the previous section in order to compute the GEV,
we write the leading order (\ie dropping $\mathcal{O}(t^{0}$) terms)
of our contour integral as
\be
{-}\frac{i\kappa^4 m_a^2 m_b^3 \gamma^2}{2^9 v^2 \sqrt{-t}}\!
   \int_{\Gamma_{\rm{LS}}}\!\!
   \frac{dy\,[2y-v(1+y^2)]^{4}}{2\pi y^3(1-y^2)^2}
\exp\!\bigg({-i}\frac{1+y^{2}-2vy}{2y-v(1+y^{2})}\bs k\times\hat{\bs p}\cdot\bs a_a
             -i \frac{1+y^{2}}{2y}\bs k\times\hat{\bs p}\cdot\bs a_b\bigg) .
\ee
As already explained, $\Gamma_{\rm{LS}}$ can be chosen as a contour around zero or infinity. This inversion accounts for a parity conjugation of the amplitude, and the equivalence follows from parity invariance of the triangle diagram~\cite{Cachazo:2017jef}.
Here let us unify both descriptions by means of the change of variables
\be
z=\frac{1+y^{2}}{2y} .
\ee
Both contours around $y=\infty$ and $y=0$ are mapped to $z=\infty$.
At the same time the polynomial structure gets reduced to at most
quadratic, at the cost of introducing a branch cut in the integral.
We now have the one-loop triangle contribution as
\be
\braket{{\cal M}_\triangleleft} = -4\pi^2\frac{G^2m_a^2m_b^3}{\sqrt{-t}}
   \int_{\Gamma_{\rm{LS}}}\frac{dz}{2\pi i}
   \frac{\gamma^2 (1-vz)^4}{v^2 (z^2-1)^{3/2}}
   \exp\!\bigg({-i}\frac{z-v}{1-vz}\bs k\times\hat{\bs p}\cdot\bs a_a
                -i z\bs k\times\hat{\bs p}\cdot\bs a_b\bigg) ,
\label{LSgev}
\ee
which now incorporates the second helicity assignment for the exchanged gravitons.
We have also inserted a factor of $-4$
to account for the HCL difference between a triangle integral
and its leading singularity.
Note that the branch cut singularity is induced by the massive propagators inside the
Compton amplitude and does not lead to classical contributions.
The essential singularity at $z=1/v$ is induced by
the unphysical pole $p_1\cdot\varepsilon_4$ in the exponential
expansion. We take the contour around infinity to be $\Gamma_{\rm{LS}}=\{|z|=R\}$ for some large but finite radius, $R>1/v$, for reasons we will explain in a moment.
Then the contribution to the scattering angle~\eqref{eq:angle} reads
\begin{align}
\theta_\triangleleft & = \pi G^2 E \frac{m_b}{2v^4} \frac{\doe}{\doe b}
\int_{\Gamma_{\rm{LS}}}\!\frac{dz}{2\pi i}\frac{(1-vz)^4}{(z^2-1)^{3/2}}
\int\!\!\frac{d^2\bs k}{2\pi|\bs k|}
\exp\!\Big( i\bs k\cdot\!\Big[ \bs b - z\hat{\bs p}\times\bs a_b
                            - \frac{z-v}{1-vz}\,\hat{\bs p}\times\bs a_a\Big]
      \Big) \nn \\ & = \pi G^2 E \frac{m_b}{2v^4} \frac{\doe}{\doe b}
\int_{\Gamma_{\rm{LS}}}\!\frac{dz}{2\pi i}\frac{(1-vz)^4}{(z^2-1)^{3/2}}
\Big|b-z a_b-\frac{z-v}{1-vz}a_a\Big|^{-1},
\label{eq:contourintegral}
\end{align}
where we have specialized to aligned spins. The total one-loop contribution to the scattering angle is $\theta_{\triangleleft}+ \theta_{\triangleright}$, where $\theta_{\triangleright}$ is obtained by exchanging $m_a \leftrightarrow m_b$ and $a_a \leftrightarrow a_b$.

Let us now discuss the choice of contour $\Gamma_{\rm{LS}}$ in
\bse
\be
\int_{\Gamma_{\rm{LS}}}\!\frac{dz}{2\pi i}\frac{(1-vz)^4}{(z^2-1)^{3/2}}
\Big|b-z a_b-\frac{z-v}{1-vz}a_a\Big|^{-1}\!=
\frac{1}{va}
\int_{\Gamma_{\rm{LS}}}\!
\frac{dz}{2\pi i}\frac{(vz-1)^5}{(z^2-1)^{3/2}(z-z_+)(z-z_-)} ,
\label{main}
\ee
where
\be
z_+ + z_- = \frac{bv+a_a+a_b}{va_b} , \qquad \quad
z_+ z_- = \frac{b+v a_a}{va_b} .
\ee
\ese
The root $z_+$ is distinguished from $z_{-}$ by demanding $z_+ \to \infty$ as $a_b \to 0$.
We now show that the appropriate leading singularity in the contour integral
is given by the residues at $z_+$ and $\infty$, by ensuring the consistency of the small-spin expansion.  If we were to take an expansion around $a_a,a_b\to0$ the poles at $z_+$ and $z_-$ would disappear at every order, leaving poles only at $z=\infty$ and $z=1/v$ together with the branch cut at $z\in (-1,1)$.  In that case, the leading-singularity prescription
in the integral~\eqref{LSgev} simply grabs the pole at $z=\infty$
and drops the branch cut contribution together with the pole at $z=1/v$.
The non-expanded expression \eqref{main} resums part of the contributions
from both $z=\infty$ and $z={1}/{v}$ into poles
located at $z_+$ and $z_-$, respectively.
This can be seen by noticing that  $z_+ \to \infty$ and $z_- \to 1/v$ as $a_a,a_b \rightarrow 0$.
This is the reason we consider a contour at finite radius $R>1/v$ in \eqn{LSgev}, so that, as long as $R<z_+$ as well,
 the contour integral can be evaluated from the poles at $z=\infty$ and $z=z_+$.

With this contour prescription,
evaluating the integral in \eqn{eq:contourintegral}
yields the explicit results given by \eqn{finalfinalresult}
in the introductory summary.  
Let us stress that
the formulas~\eqref{eq:contourintegral} and~\eqref{finalfinalresult}
can only be expected to be valid up to fourth order in $a_a$.
Nevertheless, they condense non-trivial information for the scattering angle
up to that order into a simple contour integral. 
We have checked that these results precisely match
the one-loop linear-in-spin
classical computation of \cite{Bini:2018ywr}, as well as the conjectural one-loop quadratic-in-spin expression given in \cite{Vines:2018gqi}, based on results from the exact quadrupolar test-black-hole limit \cite{Bini:2017pee} expanded to order $G^2$ and on next-to-next-to-leading-order post-Newtonian results \cite{Levi:2016ofk,Levi:2015ixa}.

\section{Discussion}
\label{sec:outro}

In this work we have presented a new connection between extended soft theorems and conservative classical gravitational observables, in particular for scattering of spinning black holes. This extends the approach initiated in~\cite{Cachazo:2017jef, Guevara:2017csg} to construct such quantities in an economic way through leading singularities. It also complements the  general picture regarding the extraction of classical results from on-shell methods, provided \eg in~\cite{BjerrumBohr:2005jr,Neill:2013wsa,Cheung:2018wkq}.

It is clear that a more precise definition is needed for the generalized expectation value that we used. Our construction can be thought as the average of an operator ${\cal O}$ as given by two particle states in the scattering amplitude, which is mapped to the expectation value of a classical observable ${\cal O}_{\text{cl}}= \braket{\cal O}$.
Interestingly, this matches their effective counterpart, as computed for instance in the worldline formalism, in the case where the operator ${\cal O}_{\text{cl}}$ is constant \cite{Vines:2017hyw,Damour:2017zjx}. An extension of the GEV may be needed to incorporate time dependence, such as what occurs with classical momentum deflection or spin holonomy \cite{Bini:2018ywr}. 

The natural desired extension of the leading-singularity method is the computation of higher orders, both in loops and powers of spin. Examples of higher-loop leading singularities were computed for gravitational theories in~\cite{Cachazo:2017jef}, so it would be interesting to see if these can be also applied to compute classical observables. On the other hand, extending the range of validity in powers of spin is now clearly related to the problem of understanding deeper orders in the soft expansion. More precisely, it is known that these orders depend both on the matter content and the coupling to gravity \cite{Laddha:2017ygw,Sen:2017xjn}, hence one could hope that such problem is tractable at least for matter
minimally coupled to gravity \cite{Arkani-Hamed:2017jhn},
thus describing black holes. Our methodology clearly resembles a soft bootstrap approach \cite{Rodina:2018pcb}, and it would be desirable to formally implement it via recursion relations \cite{Elvang:2018dco,Carballo-Rubio:2018bmu}.

It was already pointed out in~\cite{Vaidya:2014kza} that amplitudes for massive spin-$s$ particles lead to a classical potential for bodies with spin-induced multipoles such as black holes or neutron stars. The amplitudes match the classical potential up to the $2^{2s}$-pole level, or up to order~$S^{2s}$, where $S$ is the body's intrinsic angular momentum:
\begin{itemize}
\item a scalar particle corresponds to a monopole (with no higher multipoles);
\item a spin-1/2 particle adds {only} a dipole $\propto S$, yielding the ${\cal O}(S^1)$ spin-orbit effects which are universal (body-independent) in gravity;
\item a spin-1 particle further adds a spin-induced quadrupole $\propto S^2$, specifically matching the quadrupole of a spinning black hole when constructed with minimal coupling. Note that the quadrupole level corresponds to the order at which the soft theorem stops being universal.
\item a spin-3/2 particle adds a black-hole octupole $\propto S^3$, \etc
\end{itemize}
The complete spin-multipole series of a black hole is seemingly obtained
by taking the limit $s~\to~\infty$ for a massive spin-$s$ particle
minimally coupled to gravity.
This correlation was shown by Vaidya~\cite{Vaidya:2014kza}
with explicit calculations at leading post-Newtonian orders,
corresponding to the nonrelativistic limits of tree-level amplitudes,
up to the spin-2 or $S^4$ level.
In this paper, we have provided further evidence
that this correspondence holds, fully relativistically,
to all orders in spin at tree level,
and for at least the first few orders in spin at one-loop order.
It is, however, not yet clear why we should expect this correspondence
between classical black holes and minimally coupled quantum particles
with $s\to\infty$ and $\hbar\to0$,
and to what extent we should expect it to hold.

It was found in~\cite{He:2014bga}, by means of a BCFW argument,
that in the MHV sector of gravity amplitudes
there is also a natural exponential completion of the soft theorem.
A general statement for gravity amplitudes is however still missing.
There are a few evident problems for the naive extrapolation
of the formula~\eqref{eq:cachazostrominger} to higher orders.
As we have seen, increasing the powers of angular momentum, encoded
in the gauge-invariant combination $(k_\mu \varepsilon_\nu J_i^{\mu\nu})$,
requires decreasing the powers of the numerator $(p\cdot\varepsilon_i)$,
which generates unphysical poles.
Moreover, the first two orders enjoy gauge invariance
thanks to fundamental conservation laws corresponding to the linear
and angular momenta of the scattering particles \cite{Cachazo:2014fwa}.
Reinserting powers of $(p\cdot\varepsilon_i)$ in higher orders
would then impose additional constraints
that go beyond these conservation laws.
Therefore, when exponentiating the soft factor,
a very specific choice of the polarization vectors is required.
This is precisely what is done in~\cite{He:2014bga},
where this choice arises naturally from a BCFW deformation.
A second problem that we dealt with here is the sum over different particles,
which destroys the realization of the exponential
as an overall factor acting on ${\cal M}_{n-1}$.
We showed that in the cases of interest for computing
the scattering angle at tree level and one loop,
these two problems can be overcome
by a judicious choice of the polarization vectors.

An obvious question which arises from this construction is
whether it is possible to establish a link between BMS symmetries
studied at null/spatial infinity~\cite{Strominger:2013jfa,He:2014laa,
Kapec:2014opa,Dumitrescu:2015fej,Campiglia:2016hvg,Strominger:2017zoo}
(or at the black hole horizon~\cite{Hawking:2016msc,Penna:2018gfx})
and classical observables arising from massive amplitudes.
The natural candidate for such a connection is
radiative effects~\cite{Porto:2016pyg,Goldberger:2016iau,Goldberger:2017frp,
Luna:2017dtq,Shen:2018ebu},
as explored in~\cite{Laddha:2018rle} from the point of view of soft theorems.
Finally, it would be also interesting to see a link between
the exponentiation presented here and the exponentiation of IR divergences~\cite{Weinberg:1965nx,Frenkel:1976bj,Bern:1995ix,Ciafaloni:2015xsr,Strominger:2017zoo}
that has been known in QED
for a long time.
The latter one has recently appeared in the computation of tail effects
from the EFT perspective~\cite{Goldberger:2009qd,Porto:2016pyg,Levi:2018nxp}.

\begin{acknowledgments}

We would like to thank Nima Arkani-Hamed, Fabián Bautista, Freddy Cachazo, Yu-tin Huang, Ben Maybee, Matin Mojaza, Donal O'Connell, and Jan Steinhoff for useful discussions. We are very grateful to Yu-tin Huang in particular for clarifying some aspects of the gravitational and gauge couplings of massive particles in private correspondence. We are grateful to the organizers of the workshop ``QCD Meets Gravity IV'', where this work was completed. AG thanks kind hospitality from the Albert Einstein Institute, where this work was initiated, and CONICYT project 21151647 for financial support. Research at Perimeter Institute is supported by the Government of Canada through the Department of Innovation, Science and Economic Development Canada and by the Province of Ontario through the Ministry of Research, Innovation and Science.
AO has received funding from the European Union's Horizon 2020 research and innovation programme under the Marie Sklodowska-Curie grant agreement 746138.

\end{acknowledgments}

\appendix
\section{Three-point amplitude with spin-1 matter}
\label{app:spin1amp3pt}

Here we compute the three-point amplitude~\eqref{eq:spin1amp3pt}
starting from the massive spin-1 Lagrangian
\begin{equation}
{\cal L} = -\frac{1}{4} F_{\mu\nu}F^{\mu\nu} + \frac{m^2}{2} A_\mu A^\mu ,
\label{eq:proca}
\end{equation}
where $F^{\mu\nu}=\partial^\mu A^\nu-\partial^\nu A^\mu$.
In order to compute the minimal cubic vertex to gravity,
one needs to the extract the energy-momentum tensor sourced by this field.
In principle, this can be done by covariantizing this action,
\ie by promoting $\partial_{\mu}\rightarrow\nabla_{\mu}$,
and then inspecting the metric variation, $T_{\mu\nu}
=\frac{2}{\sqrt{-g}}\frac{\partial(\sqrt{-g}\mathcal{L})}{\partial g^{\mu\nu}}$.
Let us, however, take an alternative route
of computing the energy-momentum tensor directly in flat space.
The reason is that this procedure will explicitly identify
the contribution of the intrinsic angular momentum of the particle.

A textbook application of Noether's theorem for translations yields
the following tensor
\be
T_N^{\mu\nu}
 =-F^{\mu\sigma} \partial^\nu A_\sigma
 - \eta^{\mu\nu} {\cal L} \qquad \Rightarrow \qquad
\partial_\mu T_N^{\mu\nu} = 0 .
\label{eq:tuvna}
\ee
Its contraction with an on-shell graviton,
$\varepsilon_{\mu\nu} T_N^{\mu\nu}$,
fails to give the correct three-point amplitude,
as opposed to the one obtained from covariantization.
The reason is that $T_N^{\mu\nu}$ lacks symmetry
in its indices (notice \eg $\partial_\nu T_N^{\mu\nu}\neq0$),
therefore its orbital angular momentum
$L^{\lambda\,\mu\nu} = x^\mu T_N^{\lambda\nu} - x^\nu T_N^{\lambda\mu}$
is not conserved.
Let us fix that by generalizing $T_N^{\mu\nu}$ to a larger class of tensors
that are all conserved due to \eqn{eq:tuvna}:
\be
T^{\mu\nu} = T_N^{\mu\nu} + \partial_\lambda B^{\lambda\mu\,\nu} , \qquad \quad B^{\lambda\mu\,\nu} = -B^{\mu\lambda\,\nu} \qquad \Rightarrow \qquad
\partial_\mu T^{\mu\nu} = 0 ,
\ee
where the Belinfante tensor $B^{\mu\nu\rho}$
\cite{1940Phy.....7..449B,Rosenfeld:1940} may be adjusted
to yield a symmetric energy-momentum tensor matching
the gravitational one.
To do that, we apply Noether's theorem to Lorentz transformations.
The conservation of the total angular momentum
$L^{\lambda\,\mu\nu}+S^{\lambda\,\mu\nu}$ then implies
\be
T_N^{\mu\nu}-T_N^{\nu\mu} =-\partial_\lambda S^{\lambda\,\mu\nu} , \qquad \quad
S^{\lambda\,\mu\nu}
 = -i\frac{\partial{\cal L}}{\partial(\partial_\lambda A^\sigma)}
    \Sigma^{\mu\nu,\sigma}_{~~~~~\tau} A^\tau
= i F^{\lambda\sigma} \Sigma^{\mu\nu}_{\sigma\tau} A^\tau .
\ee
Here $\Sigma_{\mu\nu}$ are the Lorentz generators
$\Sigma^{\mu\nu,\sigma}_{~~~~~\tau}
=i[\eta^{\mu\sigma} \delta^\nu_\tau - \eta^{\nu\sigma} \delta^\mu_\tau]$
that will help us identify the spin contribution
inside the three-point amplitude.
Imposing that the corrected tensor $T^{\mu\nu}$ be symmetric
now yields the condition
$\partial_{\lambda}B^{\lambda[\mu\,\nu]}
=\frac{1}{2}\partial_{\lambda}S^{\lambda\,\mu\nu}$,
which is solved by 
\be
B^{\lambda\mu\,\nu} = \frac{1}{2} \big[ S^{\lambda\,\mu\nu} + S^{\mu\,\nu\lambda} - S^{\nu\,\lambda\mu} \big] .
\ee

Contracting the resulting energy-momentum tensor
with a traceless symmetric graviton~$h_{\mu\nu}$ and integrating by parts,
we obtain the gravitational interaction vertex
\beal
-h_{\mu\nu} T^{\mu\nu}
= h_{\mu\nu} F^{\mu\sigma} \partial^\nu A_\sigma
- i (\partial_\lambda h_{\mu\nu}) F^{\nu\sigma} \Sigma^{\lambda\mu}_{\sigma\tau}  A^\tau ,
\eeal
where we suppress the coupling-constant factor $\kappa/2$.
Its momentum-space version in the scattering amplitude gives the following contributions:
\begin{subequations} \begin{align}
\label{spin1fourier1}
   h_{\mu\nu} F^{\mu\sigma} \partial^\nu A_\sigma &\,\rightarrow\,
   -(p_2\!\cdot\varepsilon_3)
   \big[ (p_1\!\cdot\varepsilon_3)(\varepsilon_1\!\cdot\varepsilon_2)
       - (p_1\!\cdot\varepsilon_2)(\varepsilon_1\!\cdot\varepsilon_3)
   \big] + (1 \leftrightarrow 2) , \\ \!\!\!
\label{spin1fourier2}
  -i (\partial_\mu h_{\nu\rho}) F^{\rho\sigma}
   \Sigma^{\mu\nu}_{\sigma\tau}  A^\tau &\,\rightarrow\,
   i p_{3\mu} \varepsilon_{3\nu}
   \big[ (p_1\!\cdot\varepsilon_3)
         (\varepsilon_1\!\cdot\!\Sigma^{\mu\nu}\!\cdot\!\varepsilon_2)
       - (\varepsilon_1\!\cdot\varepsilon_3)
         (p_1\!\cdot\!\Sigma^{\mu\nu}\!\cdot\!\varepsilon_2) \big]
       + (1 \leftrightarrow 2) . \!\!
\end{align} \label{spin1fourier}%
\end{subequations}
where the transverse polarization vectors $\varepsilon_1$ and $\varepsilon_2$
correspond to the massive spin-1 matter
and two copies of $\varepsilon_3$ belong to the massless graviton.
Putting the above terms together and using the three-point on-shell kinematic conditions $p_1 \cdot p_3 = p_2 \cdot p_3 = 0$, we obtain the amplitude
\be
    {\cal M}_3
     = 2(p_1\!\cdot\varepsilon)
       \big[ (p_1\!\cdot\varepsilon) (\varepsilon_1\!\cdot\varepsilon_2)
           - 2 p_{3\mu} \varepsilon_{3\nu}
             \varepsilon_1^{[\mu} \varepsilon_2^{\nu]} \big] ,
\label{eq:spin1amplitudethree-point}
\ee
The second term in \eqn{eq:spin1amplitudethree-point} comes from
$\varepsilon_1\!\cdot\!\Sigma^{\mu\nu}\!\cdot\!\varepsilon_2^\tau
=2i\varepsilon_{1}^{[\mu}\varepsilon_{2}^{\nu]}$,
which in \app{app:spin1tensor} we interpret as a spin expectation value,
so it can be regarded as the spin contribution to the gravitational interaction.

\section{Spin tensor for spin-1 matter}
\label{app:spin1tensor}

Here we construct the spin tensor for a massive spin-1 particle
for the three-particle kinematics of \sec{sec:spin1matter}.
The starting point is the one-particle expectation value
of the angular-momentum operator in the quantum-mechanical sense:
\be
    S_p^{\mu\nu}
     = \frac{\braket{p|\Sigma^{\mu\nu}|p}}{\braket{p|p}}
     = \frac{ \varepsilon_{p\:\!\sigma}^*
              \Sigma^{\mu\nu,\sigma}_{~~~~\;\tau} \varepsilon_p^\tau }
            { \varepsilon_p^*\!\cdot\varepsilon_p }
     = 2i \varepsilon_p^{*[\mu} \varepsilon_p^{\nu]}, \qquad \quad
    \Sigma^{\mu\nu,\sigma}_{~~~~~\tau}
     = i[\eta^{\mu\sigma} \delta^\nu_\tau - \eta^{\nu\sigma} \delta^\mu_\tau] ,
\label{eq:spin1particle}
\ee
where for now we suppress the spin-projection/little-group
labels of the states.
We also used the Lorentz generators $\Sigma^{\mu\nu}$
in the vector representation.
Due to the transversality of the both massive polarization vectors,
$p\cdot\varepsilon_p = 0$, this spin tensor immediately satisfies
the SSC~\eqref{eq:ssc}.

Now a natural way to extend \eqn{eq:spin1particle}
to the case of two different states
(one incoming with momentum $p_1$ and one outgoing with $p_2$)
is to introduce a generalized expectation value
such that it gives one for a unit operator:
\be
    S_{12}^{\mu\nu} = \frac{\braket{2|\Sigma^{\mu\nu}|1}}{\braket{2|1}}
     = \frac{ \varepsilon_{2\:\!\sigma}^*
              \Sigma^{\mu\nu,\sigma}_{~~~~\;\tau} \varepsilon_1^\tau
            }{ \varepsilon_2^*\cdot \varepsilon_1 }
     = \frac{ 2i \varepsilon_2^{*[\mu} \varepsilon_1^{\nu]} }
            { \varepsilon_2^*\cdot \varepsilon_1 } .
\ee
Since in \sec{sec:multipole} we consider all momenta incoming,
we suppress the conjugation sign\footnote{The conjugation rule between
the incoming and outgoing states in the massive spinor-helicity formalism
amounts to lowering and raising the little-group indices,
as indicated by the completeness relation in \eqn{eq:polvectorsmassive}.
For instance, in the helicity basis~\cite{Arkani-Hamed:2017jhn,Ochirov:2018uyq}
of spinors for a massive momentum
$p^\mu = (E,\;\!\vec{\!p}\;\!) = (E,P\hat{p})$,
the one-particle spin quantization is explicitly
\be
   m \braket{a^\mu}_p^{ab}
    = \frac{1}{2m} \epsilon^{\mu\nu\lambda\rho}
      ( \varepsilon_{p\:\!ab}\cdot\Sigma_{\nu\lambda}\cdot
        \varepsilon_p^{ab} ) p_\rho
    = \left\{
      \begin{aligned}
      s_p^\mu &, \quad a=b=1 , \\
      0 &, \quad a+b=3 , \\
     -s_p^\mu &, \quad a=b=2 ,
      \end{aligned} \right. \qquad \quad
   \begin{aligned}
   p^\mu & = (E,\;\!\vec{\!p}\;\!) = (E,P\hat{p}) , \\
   s_p^\mu & = \frac{1}{m} (P,E\hat{p}) .
   \end{aligned}
\ee
}
and rewrite the above as
\be
    S_{12}^{\mu\nu} = -2i\varepsilon_1^{[\mu} \varepsilon_2^{\nu]}
                      /(\varepsilon_1\!\cdot\varepsilon_2) ,
\label{eq:spin1tensornaive}
\ee
which is the (normalized) angular momentum contribution obtained
in \app{app:spin1amp3pt} from Noether's theorem.
Now in a classical computation~\cite{Vines:2017hyw}
it is desirable to consider a spin tensor that 
satisfies the spin supplementary condition~\eqref{eq:ssc}.
Although \eqn{eq:spin1tensornaive} is a legitimate definition,
it does not satisy the covariant SSC~\eqref{eq:ssc}
with respect to the average momentum
$p=(p_1-p_2)/2$ of the massive particle before and after graviton emission:
\be
p_\mu S_{12}^{\mu\nu} = -\frac{i}{2}
\big( (k\cdot\varepsilon_2) \varepsilon_1^\nu
    + (k\cdot\varepsilon_1) \varepsilon_2^\nu \big)
    / (\varepsilon_1\!\cdot \varepsilon_2) \neq 0 ,
\ee
where $k=-p_1-p_2$ is the momentum transfer.
However, the spin tensor is intrinsically ambiguous,
as the separation between the orbital and intrinsic pieces
of the total angular momentum is relativistically frame-dependent.
In a classical setting, for instance, the reference point
for the intrinsic angular momentum of a spatially extended body
(as opposed to its overall orbital momentum about the origin)
is at its center of mass,
but it gets shifted by a frame change~(see \eg \cite{Steinhoff:2015ksa}).
This ambiguity allows the spin tensor to be transformed
as $S^{\mu\nu}\rightarrow S^{\mu\nu}+p^{[\mu}r^{\nu]}$,
where the difference $p^{[\mu}r^{\nu]}$ for some vector $r^\nu$ accounts
for the relative shift between $S^{\mu\nu}$
and $L^{\mu\nu} \sim p^{[\mu} \partial/\partial p_{\nu]}$.
Adjusting $r^\nu$ to accommodate for the SSC~\eqref{eq:ssc},
we obtain
\be
   S^{\mu\nu} = S_{12}^{\mu\nu}
    + \frac{2}{m^{2}} p_\lambda S_{12}^{\lambda[\mu} p^{\nu]}
    =-\frac{i}{\varepsilon_1\!\cdot \varepsilon_2} \bigg\{
      2\varepsilon_1^{[\mu} \varepsilon_2^{\nu]}
    - \frac{1}{m^2} p^{[\mu}
      \big( (k\cdot\varepsilon_2) \varepsilon_1
          + (k\cdot\varepsilon_1) \varepsilon_2 \big)^{\nu]} \bigg\} ,
\ee
where we have used that $p^2 = m^2$ for a null momentum transfer $k$.
Finally, we note that in the classical limit $k\rightarrow0$
we retrieve the spin tensor~\eqref{eq:spin1tensornaive}
as the covariant-SSC one.

\section{Angular-momentum operator}
\label{app:angularmomentum}

Here we consider the total angular momentum
\be\!
J_{\mu\nu} = L_{\mu\nu} + S_{\mu\nu} , \qquad \quad
L_{\mu\nu}^\text{pos.} = 2ix_{[\mu} \frac{\partial~}{\partial x^{\nu]}}
\ee
in terms of the spinor-helicity variables.
The starting point is the momentum-space form
of the orbital piece
\be
L_{\mu\nu} = 2ip_{[\mu} \frac{\partial~}{\partial p^{\nu]}}
 =  p_\sigma \Sigma_{\mu\nu,~\tau}^{~~~\sigma}
    \frac{\partial~}{\partial p_\tau} ,
\label{eq:orbitalmomentum}
\ee
in which we encounter the Lorentz generators $\Sigma^{\mu\nu}$ again.
Since $\Sigma_{\mu\nu,\sigma\tau}$ is antisymmetric in both pairs of indices,
we notice the subtle difference in signs between the actions
of the differential and algebraic operators,
$L_{\mu\nu} p^\rho = -\Sigma_{\mu\nu,~\sigma}^{~~~\rho} p^\sigma$,
also valid for $J_{\mu\nu}$ below.

\subsubsection*{Massless Case}

Let us warm up with the case of
a massless $k^\mu = \bra{k}\sigma^\mu|k]/2$.
The spinorial version of the angular momentum~\eqref{eq:orbitalmomentum}
is~\cite{Witten:2003nn}
\be
J^{\mu\nu}
= \bigg[ \la^\alpha \sigma_{~~~\alpha}^{\mu\nu,~\,\beta}
         \frac{\partial~}{\partial \la^\beta}
       + \lb_{\dot{\alpha}}
         \bar{\sigma}^{\mu\nu,\dot{\alpha}}_{~~~~\,\dot{\beta}}
         \frac{\partial~}{\partial \lb_{\dot{\beta}}} \bigg] ,
\label{eq:orbitalspinormassless}
\ee
where the matrices
\be
   \sigma_{~~~\alpha}^{\mu\nu,~\,\beta} = \frac{i}{4}
      \big( \sigma^{\mu}_{\alpha \dot{\gamma}}
            \bar{\sigma}^{\nu,\dot{\gamma} \beta}
          - \sigma^{\nu}_{\alpha \dot{\gamma}}
            \bar{\sigma}^{\mu,\dot{\gamma} \beta}
      \big) , \qquad \quad
   \bar{\sigma}^{\mu\nu,\dot{\alpha}}_{~~~~\,\dot{\beta}} = \frac{i}{4}
      \big( \bar{\sigma}^{\mu,\dot{\alpha} \gamma}
            \sigma^{\nu}_{\gamma \dot{\beta}}  
          - \bar{\sigma}^{\nu,\dot{\alpha} \gamma}
            \sigma^{\mu}_{\gamma \dot{\beta}}
      \big)
\label{eq:antisymsigma}
\ee
are the left-handed and right-handed representations
of the Lorentz-group algebra.
Note that the spinor map
$\{\la_\alpha,\lb_{\dot{\alpha}}\} \rightarrow k^{\mu}$
is not invertible for massless particles,
but we can still use the chain rule
\be
\frac{\partial~}{\partial \la^\alpha}
= \frac{\partial k^\mu}{\partial \la^\alpha}
  \frac{\partial~}{\partial k^\mu}
= \frac{1}{2} \sigma^\mu_{\alpha\dot{\beta}} \lb^{\dot{\beta}}
  \frac{\partial~}{\partial k^\mu} , \qquad \quad
\frac{\partial~}{\partial \lb_{\dot{\alpha}}}
= \frac{1}{2} \bar{\sigma}^{\mu,\dot{\alpha}\beta} \la_\beta
  \frac{\partial~}{\partial k^\mu}
\ee
to check the consistency
between \eqns{eq:orbitalmomentum}{eq:orbitalspinormassless}.
Namely, the action of spinorial generator on a function of momentum $k^\mu$
coincides with that of the vectorial one.

The generator~\eqref{eq:orbitalspinormassless},
which can be more concisely written in spinor indices as
\be
J_{\alpha\dot{\alpha},\beta\dot{\beta}}
= \sigma^\mu_{\alpha\dot{\alpha}}
  \sigma^\nu_{\beta\dot{\beta}} J_{\mu\nu}
= 2i\bigg[ \la_{(\alpha} \frac{\partial~}{\partial \la^{\beta)}}
           \epsilon_{\dot{\alpha}\dot{\beta}}
         + \epsilon_{\alpha\beta} \lb_{(\dot{\alpha}}
           \frac{\partial~}{\partial \lb^{\dot{\beta})}} \bigg] ,
\label{eq:orbitalspinormassless2}
\ee
has more information than its momentum-space counterpart,
as it cares about the helicity of the massless particle.
For instance, when we write the polarization tensors
in terms of spinor-helicity variables,
\be
   \varepsilon^+_{\alpha\dot{\alpha}} = \sqrt{2}\;\!
      \frac{\ket{r}_\alpha [k|_{\dot{\alpha}}}{\braket{r\;\!k}} , \qquad \quad
   \varepsilon^-_{\alpha\dot{\alpha}} = -\sqrt{2}\;\!
      \frac{\ket{k}_\alpha [r|_{\dot{\alpha}}}{[r\;\!k]} ,
\label{eq:polvectors2}
\ee
we do not regard them as functions of $k^\mu$
but rather of its spinors $\la_\alpha$ and $\lb_{\dot{\alpha}}$.
Of course,
an integer spin should not by itself depend on the auxiliary spinors.
Fortunately, we can show that the action
of the differential operator~\eqref{eq:orbitalspinormassless2}
is precisely that of the algebraic generator~$\Sigma_{\mu\nu}$,
which constitutes the intrinsic angular momentum
\be
(\varepsilon S^{\mu\nu})_\tau
= \varepsilon_\sigma \Sigma^{\mu\nu,\sigma}_{~~~~~\tau}
= 2i \varepsilon^{[\mu} \delta^{\nu]}_\tau
\qquad \Rightarrow \qquad
(\varepsilon S_{\alpha\dot{\alpha},\beta\dot{\beta}})_{\gamma\dot{\gamma}}
= 2i[ \varepsilon_{\alpha\dot{\alpha}}
      \epsilon_{\beta\gamma} \epsilon_{\dot{\beta}\dot{\gamma}}
    - \epsilon_{\alpha\gamma} \epsilon_{\dot{\alpha}\dot{\gamma}}
      \varepsilon_{\beta\dot{\beta}} ] .
\ee
Specializing to the negative-helicity case for concreteness,
we indeed find
\beal
J_{\alpha\dot{\alpha},\beta\dot{\beta}} \varepsilon^-_{\gamma\dot{\gamma}}
= ( \varepsilon^-
    S_{\alpha\dot{\alpha},\beta\dot{\beta}} )_{\gamma\dot{\gamma}}
  + \frac{2\sqrt{2}i}{[q\;\!k]^2} \epsilon_{\alpha\beta}
    [q|_{\dot{\alpha}} [q|_{\dot{\beta}} \ket{k}_\gamma [k|_{\dot{\gamma}} .
\eeal 
Here the last term is a gauge contribution
explicitly proportional to $k_{\gamma\dot{\gamma}}$,
so it can be discarded in a physical amplitude.

Therefore, we conclude that
the spinorial differential operator~\eqref{eq:orbitalspinormassless2}
incorporates both the orbital and intrinsic contributions,
so it is the total angular-momentum operator.

\subsubsection*{Massive Case}

It is direct to generalize the above discussion
to massive momenta $p^\mu=\bra{p^a}\sigma^\mu|p_a]/2$
\cite{Conde:2016izb}.
The angular-momentum operator
in the space of massive spinors $\{\la_\alpha^a,\lb_{\dot{\beta}}^b\}$
is given by
\be
J^{\mu\nu}
= \bigg[ \la^{\alpha a} \sigma_{~~~\alpha}^{\mu\nu,~\,\beta}
         \frac{\partial~}{\partial \la^{\beta a}}
       + \lb_{\dot{\alpha}}^{\,a}
         \bar{\sigma}^{\mu\nu,\dot{\alpha}}_{~~~~\,\dot{\beta}}
         \frac{\partial~}{\partial \lb_{\dot{\beta}}^a} \bigg] , \qquad \quad
J_{\alpha\dot{\alpha},\beta\dot{\beta}}
= 2i\bigg[ \la_{(\alpha}^{~a} \frac{\partial~}{\partial \la^{\beta)a}}
           \epsilon_{\dot{\alpha}\dot{\beta}}
         + \epsilon_{\alpha\beta} \lb_{(\dot{\alpha}}^{~a}
           \frac{\partial~}{\partial \lb^{\dot{\beta})a}} \bigg] .
\label{eq:orbitalspinormassive}
\ee
This operator is by construction invariant under
the little group ${\rm SU}(2)$.
Using the chain rule
\be
\frac{\partial~}{\partial \la^{\alpha a}}
= \frac{\partial p^\mu}{\partial \la^{\alpha a}}
  \frac{\partial~}{\partial p^\mu}
= \frac{1}{2} \sigma^\mu_{\alpha\dot{\beta}} \lb^{\dot{\beta}}_a
  \frac{\partial~}{\partial p^\mu} , \qquad \quad
\frac{\partial~}{\partial \lb_{\dot{\alpha}}^a}
=-\frac{1}{2} \bar{\sigma}^{\mu,\dot{\alpha}\beta} \la_{\beta a}
  \frac{\partial~}{\partial p^\mu} ,
\ee
it is again easy to check that the action on a function of
$p_{\alpha\dot{\beta}}=\la_\alpha^a \epsilon_{ab} \lb_{\dot{\beta}}^b$
is the same as that of \eqn{eq:orbitalmomentum}.
Finally, the action on polarization tensors
can be tested to be a Lorentz transformation.
The spin-$s$ tensors are parametrized
in terms of massive spinor-helicity variables as
\be
\varepsilon_{\alpha_1\dot{\alpha}_1\ldots
             \alpha_s\dot{\alpha}_s}^{a_1\ldots a_{2s}}
= \frac{2^{s/2}}{m^s} \la_{\alpha_1}^{(a_1} \lb_{\dot{\alpha}_1}^{a_2}
  \cdots \la_{\alpha_s}^{a_{2s-1}} \lb_{\dot{\alpha}_s}^{a_{2s})} .
\label{eq:poltensors}
\ee
with an obvious extension by an additional factor
of Dirac spinor~\cite{Arkani-Hamed:2017jhn,Ochirov:2018uyq}
for half-integer spins.
Indeed, since $J^{\mu\nu}$ is a first-order differential operator,
it distributes when acting on $\varepsilon^{a_{1}\cdots a_{2s}}$
and naturally expands into the left- and right-handed Lorentz generators:
\beal
J^{\mu\nu} \varepsilon_{\alpha_1\dot{\alpha}_1\ldots
                        \alpha_s\dot{\alpha}_s}^{a_1\ldots a_{2s}} =
    \frac{2^{s/2}}{m^s} \bigg\{
    \Big[ \epsilon_{\alpha_1 \beta} \big( \la^{\alpha(a_1} \sigma_{~~~\alpha}^{\mu\nu,~\,\beta} \big) \Big]
    \lb_{\dot{\alpha}_1}^{a_2} & \cdots
    \la_{\alpha_s}^{a_{2s-1}} \lb_{\dot{\alpha}_s}^{a_{2s})} \\
  + \la_{\alpha_1}^{(a_1}
    \Big[ \lb_{\dot{\alpha}}^{a_2} \bar{\sigma}^{\mu\nu,\dot{\alpha}}_{~~~~\,\dot{\alpha_2}}
    \Big] & \cdots
    \la_{\alpha_s}^{a_{2s-1}} \lb_{\dot{\alpha}_s}^{a_{2s})}
  + \ldots \bigg\} .
\eeal

\bibliographystyle{JHEP}
\bibliography{references}

\end{document}